\documentclass[sigconf]{acmart}
\usepackage{multirow}
\usepackage{subfigure}
\usepackage{amsmath}
\usepackage{amsthm}
\usepackage{amsfonts} 
\usepackage{algorithm}
\usepackage{algorithmic}
\usepackage{booktabs}
\usepackage{graphicx}
\usepackage{url}
\usepackage{textcomp}
\usepackage{xcolor}
\usepackage{subfigure} 
\usepackage{float}
\usepackage{multirow}
\usepackage{multicol}
\usepackage{color}
\usepackage{bm}
\usepackage{bbm}
\usepackage{enumitem}
\usepackage{lineno}
\usepackage[normalem]{ulem} 
\usepackage{CJKutf8}
\newtheorem{theorem}{Theorem}[section]

\newtheorem{prop}[theorem]{Proposition}

\usepackage{pifont}
\def \mL{\mathcal L}

\def \R{\mathbb R}
\newcommand{\agg}{\text{AGG}}
\newcommand{\update}{\text{UPD}}
\newcommand{\ms}[2]{$#1_{(#2)}$}
\newcommand{\bms}[2]{$\mathbf{#1}_{(#2)}$}
\newcommand{\sbms}[2]{$\underline{#1}_{(#2)}$}

\def\etal{et al.\;}

\newcommand{\be}{\mathbf{e}}

\newcommand{\bh}{\mathbf{h}}

\newcommand{\bH}{\mathbf{H}}
\newcommand{\bW}{\mathbf{W}}

\newcommand{\bA}{\mathbf{A}}
\newcommand{\bQ}{\mathbf{Q}}
\newcommand{\bK}{\mathbf{K}}
\newcommand{\bV}{\mathbf{V}}

\newcommand{\cU}{\mathcal{U}}
\newcommand{\cV}{\mathcal{V}}

\newcommand{\cS}{\mathcal{S}}
\newcommand{\cQ}{\mathcal{Q}}
\newcommand{\cT}{\mathcal{T}}

\newcommand{\btheta}{\bm{\theta}}
\newcommand{\bphi}{\bm{\phi}}

\usepackage{array}
\newcolumntype{L}[1]{>{\raggedright\let\newline\\\arraybackslash\hspace{0pt}}m{#1}}
\newcolumntype{C}[1]{>{\centering\let\newline  \\\arraybackslash\hspace{0pt}}m{#1}}
\newcolumntype{R}[1]{>{\raggedleft\let\newline \\\arraybackslash\hspace{0pt}}m{#1}}

\newcommand{\TheName}[0]{EmerG}

\AtBeginDocument{%
	}

\copyrightyear{2024}
\acmYear{2024}
\setcopyright{acmlicensed}\acmConference[KDD '24]{Proceedings of the 30th ACM SIGKDD Conference on Knowledge Discovery and Data Mining}{August 25--29, 2024}{Barcelona, Spain}
\acmBooktitle{Proceedings of the 30th ACM SIGKDD Conference on Knowledge Discovery and Data Mining (KDD '24), August 25--29, 2024, Barcelona, Spain}
\acmDOI{10.1145/3637528.3671784}
\acmISBN{979-8-4007-0490-1/24/08}

\author{Yaqing Wang}
\affiliation{%
	\institution{Baidu Inc.}
	\department{Baidu Research}
	\city{Beijing}
	\country{China}}
\email{wangyaqing01@baidu.com}
\author{Hongming Piao}
\affiliation{%
	\institution{City University of Hong Kong}
	\department{Department of Computer Science}
	\city{Hong Kong SAR}
	\country{China}}
\email{hpiao6-c@my.cityu.edu.hk}
\author{Daxiang Dong}
\affiliation{%
	\institution{Baidu Inc.}
	\department{Baidu AI Cloud}
	\city{Beijing}
	\country{China}}
\email{dongdaxiang@baidu.com}
\author{Quanming Yao}
\affiliation{%
	\institution{Tsinghua University}
	\department{Department of Electronic Engineering}
	\city{Beijing}
	\country{China}}
\email{qyaoaa@tsinghua.edu.cn}
\author{Jingbo Zhou}
\affiliation{%
	\institution{Baidu Inc.}
	\department{Baidu Research}
	\city{Beijing}
	\country{China}}
\email{zhoujingbo@baidu.com}

\begin{document}
\title[\TheName{}]{Warming Up Cold-Start CTR Prediction \\
	by Learning Item-Specific Feature Interactions}

\begin{abstract}
In recommendation systems, new items are continuously introduced, initially lacking interaction records but gradually accumulating them over time. 
Accurately predicting the click-through rate (CTR) for these items is crucial for enhancing both revenue and user experience. While existing methods focus on enhancing item ID embeddings for new items within general CTR models, they tend to adopt a global feature interaction approach, often overshadowing new items with sparse data by those with abundant interactions. Addressing this, our work introduces EmerG, a novel approach that warms up cold-start CTR prediction by learning item-specific feature interaction patterns. EmerG utilizes hypernetworks to generate an item-specific feature graph based on item characteristics, which is then processed by a Graph Neural Network (GNN). This GNN is specially tailored to provably capture feature interactions at any order through a customized message passing mechanism. 
We further design a meta learning strategy that optimizes parameters of hypernetworks and GNN across various item CTR prediction tasks, while only adjusting a minimal set of item-specific parameters within each task. This strategy effectively reduces the risk of overfitting when dealing with limited data. 
Extensive experiments on benchmark datasets validate that EmerG consistently performs the best given no, a few and sufficient instances of new items. 
\end{abstract}


\maketitle

\section{Introduction}
\label{sec:intro}

The cold-start problem presents a significant challenge in recommender systems~\cite{park2008long}, particularly evident as new items transition from having no user interactions (termed as cold-start phase) to accumulating a few initial clicks (termed as warm-up phase)  in the industry landscape. 
Deep learning models, renowned for their capability to capture complex feature interactions, have shown promise in improving click-through rate (CTR) predictions, a critical metric for assessing the likelihood of user engagement with various items (e.g., movies, commodities, music)~\cite{WD,DeepFM,bian2023feynman,wen2023efficient}. However, these models typically rely on extensive datasets to achieve optimal performance, a requirement that poses a limitation in cold-start and warm-up phases. With their substantial parameter size, these models struggle to adapt efficiently to these phases characterized by limited interaction records, thereby exacerbating the challenge of making accurate CTR predictions and updating models without incurring significant costs. 

Recent studies have focused on enhancing the initialization of item ID embeddings as a strategy to mitigate the item cold-start problem in recommender systems, which allows subsequent updates through gradient descent as interaction records become available in the warm-up phase~\cite{metaE,MWUF,CVAR,GME}. 
Then, they  leverage general CTR backbones for further processing. 
However, they overlook a crucial aspect: the distinctiveness of feature interaction patterns across different users and items. This oversight limits the ability of these models to fully capture the nuanced dynamics of user-item interactions, potentially impacting the accuracy and effectiveness of CTR predictions in scenarios where personalized recommendations are crucial.
For example, comparing high-priced luxury items with low-priced daily necessities reveals distinct interaction patterns. 
For high-priced luxury items, the interaction between the item's price and the user's income level is pivotal. Specifically, the second-order feature interaction <price, income> can be a determining factor in the user’s willingness to purchase such items. As for low-priced daily necessities, the impact of the user's income level on purchasing decisions diminishes. In these instances, other feature interactions, such as those between the user’s age and the item’s category, become relatively more important. This variation underscores the necessity of modeling item-specific feature interaction patterns. While existing works all learn a global feature interaction pattern between users and items, which overwhelms new items with a limited number of interaction records by old items with abundant interaction records. 


Recognizing the crucial role of feature interactions, 
we introduce \TheName{} to address CTR prediction of newly emerging items with incremental interaction data (from no interaction records to few and then abundant records) through the learning of item-specific feature graphs. 
Our contributions can be summarized as follows: 
\begin{itemize}[leftmargin=*]
	\item We propose a unique method that emphasizes item-specific feature interactions, addressing the challenge of new item CTR prediction by reducing the overshadowing effect of older items with extensive data. 
	Utilizing hypernetworks, we construct item-specific feature graphs with nodes as features and edges as their interactions, capturing complex interaction patterns unique to each item.
	We use a graph neural network (GNN) with a customized message passing process designed to provably capture feature interactions at any orders, which can be combined into nuanced and accurate predictions.
	\item To mitigate overfitting given limited data, we adopt a meta-learning strategy that optimizes parameters of hypernetworks and GNN  
	across different item CTR prediction tasks with a few adjustments to item-specific parameters within each task. Besides, hypernetworks and GNN learned this way are expected to generalize to each task easily. 
	\item We conduct extensive experiments on benchmark datasets and validate that \TheName{} performs the best for CTR prediction of emerging items. 
	We also evaluate the performance given more training data, and find that \TheName{} consistently performs the best. 
	Visualization of adjacency matrices which record item-specific feature graphs shows that \TheName{} can learn item-specific feature interactions properly. 
\end{itemize}

\section{Related Works}
\label{sec:related}


We briefly review four groups of methods relevant to CTR prediction of newly emerging items. 
\paragraph{A. General CTR Models.}
General CTR models are applied universally without prioritizing underrepresented items. 
They mainly focus on 
modeling complex feature interactions, which is crucial for enhancing CTR prediction accuracy~\cite{feature_interaction}. 
Historical advancements in this area show a progression  from 
simple first-order interactions captured by linear models like logistic regression \cite{lr} to second-order interactions modeled by Factorization Machines (FM) \cite{fm}, and to 
high-order interactions addressed by higher-order FMs (HOFMs)~\cite{hofm}. 
Various deep-learning models then automate the learning of these complex patterns.
Both Wide\&Deep \cite{WD} and DeepFM \cite{DeepFM}  employ hybrid architectures to handle second-order and higher interactions. 
DIN~\cite{zhou2018deep} dynamically captures user interests through an attention mechanism that adapts to varying ad features. 
AutoInt \cite{AutoInt} introduces a multi-head self-attention mechanism \cite{attention} to model high-order feature interactions. 
LorentzFM \cite{lorentzfm} explores feature interactions in hyperbolic space to minimize parameter size.
AFN \cite{AFN} converts the power of each feature into the coefficient to be learned. 
FinalMLP~\cite{mao2023finalmlp} uses 
two multi-layer perceptron (MLP) networks in parallel, and 
equips them with feature selection and interaction aggregation layers. 
FINAL~\cite{zhu2023final} introduces a factorized interaction layer for exponential growth in feature interaction learning. 
Considering that feature interactions can be conceptualized as graphs with nodes representing user and item features, 
graph neural network (GNN)~\cite{kipf2017semi} are used. 
Fi-GNN \cite{fignn} learns to generate feature graphs where the edges are established according to the similarity between feature embeddings. 
FIVES \cite{fives} learns a global feature graph where edges are established by differentiable search from large-scale CTR datasets, thus new items with limited data can be underrepresented. 
GMT~\cite{min2022neighbour} 
models the interactions among items, users, and their features into a large heterogeneous graph, then feeds the local neighborhood of the target user-item pair for prediction. 
However, these general CTR models, 
designed for extensive datasets, often struggle during the cold-start \& warm-up phases due to their substantial parameter size, which can lead to overfitting when adapted for new items. In contrast, \TheName{} introduces a specialized GNN that operates on item-specific feature graphs, generated via hypernetworks learned from diverse CTR tasks, ensuring precise modeling of feature interactions for new items with minimal parameter adjustments.


\paragraph{B. Methods for New Items without Interaction Records.} 
Several methods specifically address the cold-start phase, where new items lack interaction records, while still maintaining model performance on older, established items. 
DropoutNet~\cite{volkovs2017dropoutnet} trains neural networks with a dropout mechanism on input samples to infer missing data. 
Heater~\cite{zhu2020recommendation} employs a multi-gate mixture-of-experts approach to generate item embeddings. 
GAR~\cite{chen2022generative} adopts an adversarial training strategy between a generator and a recommender to produce new item embeddings that mimic the distribution of old embeddings, deceiving the recommender systems. 
ALDI~\cite{ALDI} learns to transfer the behavioral information of old items to new items. 
However, these methods do not consider the incorporation of incoming interaction records for new items and typically require re-training to accommodate the evolving interaction history of these items.

\paragraph{C. Methods for New Items with A Few Interaction Records.} 
In scenarios where only a few instances of new items are available, which correspond to warm-up phases, 
few-shot learning~\cite{wang2020generalizing} present a natural solution. 
Few-shot learning targets at generalizing to new tasks with a few labeled samples, which has been applied to image classification~\cite{finn2017model},  query intent recognition~\cite{wang2022recognizing} and drug discovery~\cite{wang2021property,yao2024property,wu2024pacia}. 
For CTR prediction, 
existing works  typically approach the problem as a $N$-way $K$-shot task, where each of the $N$ new items is associated with $K$ labeled instances, then  
utilize the classic gradient-based meta-learning strategy~\cite{finn2017model}. 
MeLU~\cite{melu} leverages this strategy to selectively adapt model parameters for new items through gradient descents. 
MAMO~\cite{mamo} enhances adaptation by incorporating an external memory mechanism. 
MetaHIN~\cite{lu2020meta} utilizes heterogeneous information networks to exploit the rich semantic relationships between users and items. 
PAML~\cite{wang2021preference} employs social relations to facilitate information sharing among similar users.
More recent approaches have shifted from user-specific fine-tuning via gradient descent to amortization-based methods. 
These methods directly map user interaction histories to user-specific parameters, thus modulating the main network without iterative adjustments. 
TaNP~\cite{TaNP} learns to modulate item-specific parameters based on item interaction records.  
ColdNAS~\cite{ColdNAS} employs neural architecture search to optimize the modulation function and its application within the network. 
However, these methods struggle to handle new items that lack interaction records and cannot dynamically incorporate additional interaction records of new items as they become available.

\paragraph{D. Methods for Emerging Items with Incremental Interaction Records.} 
To mirror the industry's dynamic evolution of new items, progressing from no interaction records to few and then abundant records, 
models are developed to manage these transitions smoothly. 
Existing efforts primarily enhance item ID embeddings for general CTR backbones. 
MetaE \cite{metaE} employs gradient-based meta-learning to train an embedding generator. 
MWUF \cite{MWUF} transforms unstable item ID embeddings into stable ones using meta networks. 
CVAR \cite{CVAR} decodes new item ID embeddings from a distribution over item side information, circumventing additional data processing. 
GME \cite{GME} leverages information from neighboring old items for new item ID embedding generation. 
However, these methods generally optimize initial item ID embeddings while maintaining a global feature interaction pattern, thus failing to capture the unique characteristics and interaction dynamics of these items. 
In contrast, 
our \TheName{} addresses new item CTR prediction by tailoring feature interactions to each item with the help of hypernetworks. 
By learning with item-specific feature interactions,   
the risk of overwhelming new items by old items with abundant data is alleviated
and prediction accuracy is enhanced.

%

\begin{figure*}[hbtp]
	\centering
	\includegraphics[width=0.96\textwidth]{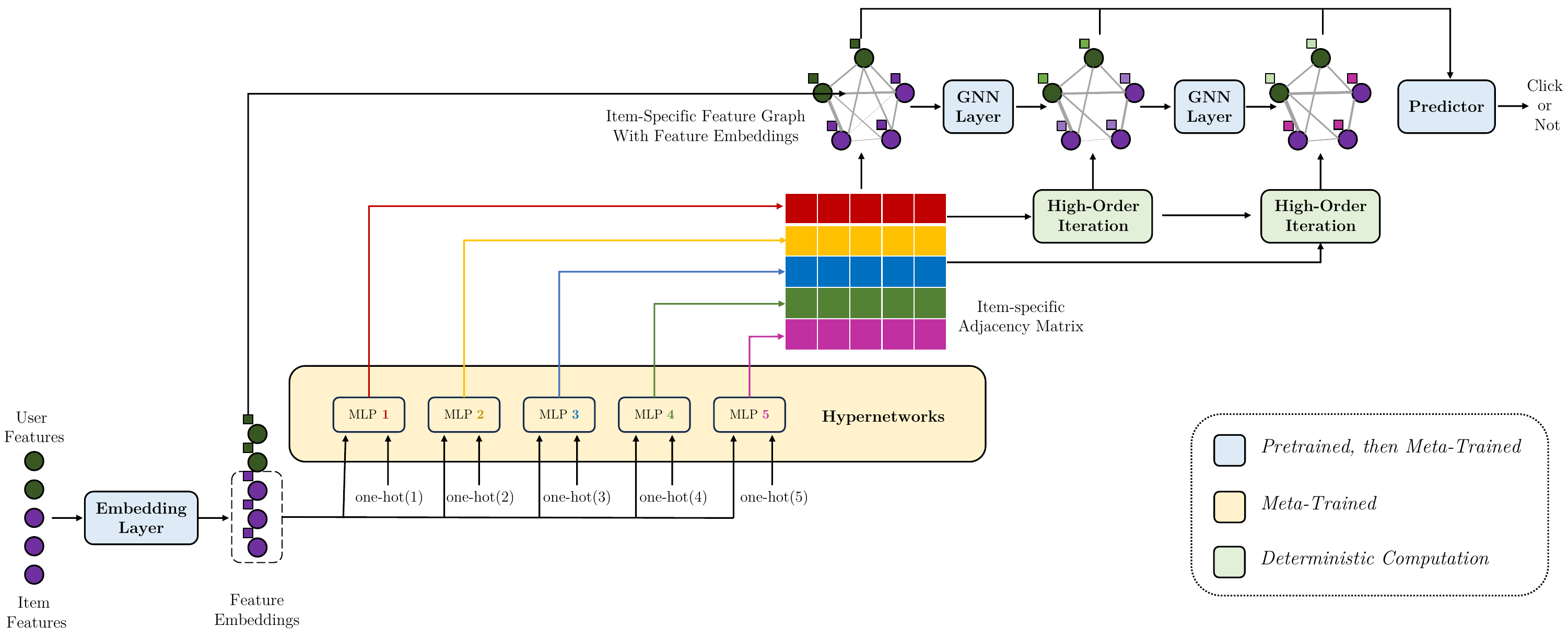}
	\vspace{-5px}
	\caption{Illustration of the proposed \TheName{}, designed to enhance CTR predictions of newly emerging items through the learning of item-specific feature interaction patterns. 
	\TheName{} uses hypernetworks to generate an initial item-specific adjacency matrix for a feature graph, with nodes representing user and item features and edges denoting their interactions, based on item feature embeddings. 
	Higher-order adjacency matrices for subsequent GNN layers are generated from the initial matrix, reducing both model complexity and storage requirements. The GNN's message passing process is tailored to capture $l$-order feature interactions at the $l-1$th layer, enabling nuanced integration of various interaction orders for accurate predictions. \TheName{} optimizes the parameters of hypernetworks and GNN across diverse CTR prediction tasks to enhance generalization, while utilizing minimal item-specific parameters to capture the uniqueness of new items, which are adaptable with the introduction of additional item instances. 
	}
	\vspace{-5px}
	\label{fig:illus}
\end{figure*}


\section{Problem Formulation}
\label{sec:problem}

Let $\cV = \{v_i\}$ denote a set of items where each item $v_i$ is associated with $N_v$  item features such as item ID, type and price.  
Similarly, let $\cU = \{u_j\}$ denote a set of users where each user $u_j$ is also associated with $N_u$ user features such as user ID, age and hometown. 
When a user $u_j$ clicks through an item $v_i$, label $y_{i,j}=1$. Otherwise,  $y_{i,j}=0$. 

During learning, 
the predictor is learned from a set of CTR prediction tasks 
$\cT^{\text{old}}=\{\cT_i\}_{t=1}^{N_t}$ sampled from old items, 
which can rapidly generalize to predict for tasks from new items that are unseen during training.  
Each task $\cT_i$ corresponds to an old item $v_i$, 
with a training set $\cS_i = \{(v_i,u_j, y_{i,j})\}_{ j = 1 }^{N_s}$ containing existing interaction histories associated with $v_i$
and a test set $\cQ_i = \{(v_i, u_{j},y_{i,j})\}_{j=1}^{N_q}$ containing interactions to predict whether $u_j$ clicks through $v_i$. 
$N_s$ and $N_q$ are the number of interactions in $\cS_i$ and $\cQ_i$ respectively. 

During testing, we 
consider CTR prediction for new items that start with no interaction records, then gradually gather a few, and eventually accumulate sufficient interaction records.  
Consider a task $\cT_k$ associated with a new item $v_k$ which is not considered during training, 
we handle three phases: 
\begin{itemize}[leftmargin=*]
	\item Cold-start phase: No training set is provided. 
	\item Warm-up phase: 
	A training set $\cS_{k}$ containing 
	a few interaction records of $v_k$ is given. There can be multiple warm-up phases where interaction records are gradually accumulated. 
	\item Common phase: 
	The training interaction records of $v_k$ are accumulated to be sufficient. 
\end{itemize}
For all three phases, the performance is evaluated on test set $\cQ_k$.

\section{The Proposed \TheName{}}
\label{sec:proposed}

Aligning with the established understanding that feature interactions are crucial, we propose \TheName{} (Figure~\ref{fig:illus}) to capture the uniqueness of items through their associated feature interaction patterns. We design two key components in \TheName{}:  
(ii) hypernetworks shared across different tasks to generate item-specific adjacent matrices encoding feature graphs; and
(i) a GNN that operates on the generated item-specific feature graphs, 
whose message passing mechanism is specially tailored to provably capture feature interactions at any order. 
As we consider cold-start \& warm-up phases, we further design a meta learning strategy that optimizes parameters of hypernetworks and GNN across various item CTR prediction tasks, while only adjusting a small set of item-specific parameters within each task. This strategy effectively reduces the risk of overfitting when dealing with limited data.



\subsection{Embedding Layer} 
Given an instance $(u,v)$, embedding layer
maps the user features of $u$ and item features of $v$ into dense vectors. 
For the $m$th feature $f_m, m\in[1,N_v+N_u]$, 
its feature embedding $\be_m$ is obtained as
\begin{align}\label{eq:emb-layer}
	\be_m= 
	\begin{cases}
		\bm{W}_{e,m} \cdot \text{one-hot}(f_m)& \text{if $f_m$ is single-valued}\\
		\sum_e{\bm{W}_{e,m} \cdot \text{multi-hot}(f_m)}& \text{if $f_m$ is multi-valued}\\
		\bm{W}_{e,m} \cdot f_m&\text{if $f_m$ is continuous}
	\end{cases}, 
\end{align}
where 
$\bm{W}_{e,m}$ represents the embedding matrix corresponding to the $m$th feature, 
$\text{one-hot}(f_m)$ represents the one-hot vector of single-valued feature $f_m$, 
and $\text{multi-hot}(f_m)$ represents the multi-hot vector of multi-valued feature  $f_m$. 

\subsection{Item-Specific Feature Graph Generation} 
\label{sec:proposed-hypernet}

We employ hypernetworks, following the strategy of Ha \etal~\cite{ha2017hypernetworks}, to generate item-specific feature graphs. Hypernetworks, small neural networks trained to generate parameters for a larger main network, present a unique challenge in their application, as their integration is highly problem-specific. In \TheName{}, hypernetworks are used to produce the initial adjacency matrix $\bA^{(1)}_i$, encoding the item-specific feature graph for the first GNN layer. 
We streamline the process by allowing subsequent GNN layers to derive their adjacency matrices from the initial $\bA^{(1)}_i$, optimizing storage efficiency without compromising the model's specificity to each item. 

Consider 
task $\cT_i$ for item $v_i$. 
For item features $f_1,\dots,f_{N_v}$, 
item feature embeddings are denoted as $\be_{1,i},\dots,\be_{N_v,i}$ respectively. 
The 
feature graph~\cite{fignn,fives} is a graph 
where each node corresponds to a feature $f_m$, and the edge between two nodes records their interaction. 
We let 
our
hypernetworks, which consists of  $N_v+N_u$ subnetworks, produce
a dense item-specific $\bar{\bA}^{(1)}_i\in\R^{(N_v+N_u)\times (N_v+N_u)}$ which encodes the feature graph to be used in the first GNN layer. 
Denote the $m$th row of $\bar{\bA}^{(1)}_i$
as$[\bar{\bA}^{(1)}_i]_{m:}$, which is calculated as: 
\begin{align}
	[\bar{\bA}^{(1)}_i]_{m:} = {\text{MLP}}_{\bW_a}( [\be_{1,i},\dots,\be_{N_v,i},\text{one-hot}(m)]), 
	\label{eq:hyper1}
\end{align}
where ${\text{MLP}}_{\bW_a}$ denotes a multi-layer perceptron (MLP) with parameter $\bW_a$. 
Then, we generate $\bar{\bA}^{(l)}_i$ as
\begin{equation}
	\label{eq:a-iter}
	\bar{\bA}^{(l)}_i = \bar{\bA}^{(l-1)}_i \cdot \bar{\bA}^{(1)}_i.
\end{equation}
This \eqref{eq:a-iter} returns $\bar{\bA}^{(l)}_i$ as the matrix product of $l$ copies of $\bar{\bA}^{(1)}_i$. Therefore,  $[\bar{\bA}^{(l)}_i]_{mn}$ records the number of $l$-hop paths from node $m$ to node $n$. 
In this way, we only need to keep one adjacency matrix (i.e., $\bar{\bA}^{(1)}_i$) for each item $v_i$ no matter how many GNN layers are used, which reduces parameter size. 

Further, 
we take the following steps to refine adjacency matrices:  
\begin{align}
	\label{eq:a-sparsify}
	\hat{\bA}^{(l)}_i &= \text{sparsify}(\text{normalize}(\bar{\bA}^{(l)}_i),K), \\
	\label{eq:symmetric}
	\tilde{\bA}^{(l)}_i &= {((\hat{\bA}^{(l)}_i)^\top + \hat{\bA}^{(l)}_i)}/{2},\\
	\label{eq:mask}
	\bA^{(l)}_i &= \text{normalize}( \text{mask}([ \tilde{\bA}^{(l-1)}_i \cdot \tilde{\bA}^{(1)}_i], \tilde{\bA}^{(l-1)}_i)),
\end{align}
where $\text{normalize}(\cdot)$ applies min-max normalization to scale all the elements of a matrix to be in the range $[0,1]$ and sets the diagonal elements of a matrix directly as 1, 
$\text{sparsify}(\cdot,K)$ keeps the top $K$ largest elements and set the rest as 0, 
and $\text{mask}(\cdot,\tilde{\bA}^{(l-1)}_i)$ sets all zero elements of $\tilde{\bA}^{(l-1)}_i$ in $\tilde{\bA}^{(l)}_i$ as zero. 
From \eqref{eq:a-sparsify} to \eqref{eq:mask}, 
we first 
sparsify the dense $\bar{\bA}^{(l)}_i$ by \eqref{eq:a-sparsify}
such that only two highly related features are connected. 
Further, because of the commutative law of $\odot$, we transform $\hat{\bA}^{(l)}_i$ into a symmetric matrix by \eqref{eq:symmetric}. 
Apart from the above-mentioned considerations, 
we expect that 
nodes disconnected in low-order feature graphs to be disconnected in high-order feature graphs. 
For example, if the message is not propagated from node $n_2$ to node $n_1$ in the $l$th GNN layer, the message of $n_2$ will be not propagated to $n_1$ in higher GNN layers. 
Thus, we apply \eqref{eq:mask} to obtain the final $\bA^{(l)}_i$.  

\subsection{Customized Message Passing Process on Item-Specific Feature Graph}
\label{sec:proposed-gnn}
Upon the learned item-specific feature graphs, we 
use a GNN with a customized message passing process designed to provably capture feature interactions at any orders, which are then explicitly combined into the final CTR predictions. 

We first describe the general mechanism of message passing. 
At the $l$th GNN layer, node embedding $\bh_{m}^{(l)}$ of feature $f_m$ is updated as 
\begin{align}
	\label{eq:gnn-update-general}
	\bh_{m}^{(l)}
	=\update^{(l)}
	\left( 
	\bh_{m}^{(l-1)},
	\agg^{(l)} 
	\left(
	\left\{
	\bh_{n}^{(l-1)} : f_n \in \mathcal{N}(f_m)
	\right\}
	\right)\right),
\end{align}
where 
$\update^{(l)}(\cdot)$ updates node embedding of $f_m$ as $\bh_{m}^{(l)}$, 
$\agg^{(l)}(\cdot)$ aggregates node embeddings of neighbor nodes, 
$\mathcal{N}(f_m)$ contains neighbor nodes of node corresponding to feature $f_m$, 
and 
$\bh^{(0)}_m=\be_m$.  
After $N_l$ layers, node embedding $\bh_{m}=\bh_{m}^{(N_l)}$ is returned as the final feature representation. 

In \TheName{}, 
we realize \eqref{eq:gnn-update-general} as 
\begin{align}
	\label{eq:gnn-update}
	\bh_{m}^{(l)}
	=
	\bh_{m}^{(l-1)}\odot
	\left[
	\sum\nolimits_{n = 1}^{N_v+N_u}{[\bA^{(l - 1)}_i]_{mn} \bW^{(l - 1)}_g\bh_{n}^{(0)}}
	\right],
\end{align}
where 
$\odot$ is the element-wise product,  
$\bA^{(l - 1)}_i$ is the final item-specific adjacency matrix obtained by \eqref{eq:mask}, 
and $\bW^{(l - 1)}_g$ is a learnable parameter. 
Unlike existing GNNs~\cite{fignn,fives,kipf2017semi} that aggregate $\bh_{m}^{(l-1)}$ with $\bh_{n}^{(l-1)}$, we aggregate $\bh_{m}^{(l-1)}$ with $\bh_{n}^{(0)}$. In this way, as Proposition \ref{prop:gnn} shows, the output of $(l-1)$th GNN layer is $l$-order feature interactions, which enables \TheName{} to explicitly model arbitrary-order feature interaction. 
\begin{prop}[Efficacy of \TheName{}.] 
	\label{prop:gnn}
	With the customized message passing process defined in \eqref{eq:gnn-update}, 
	the $(l-1)$th GNN layer captures 
	$l$-order feature interactions. 
\end{prop}
The proof is in Appendix \ref{app:proof}. 
One may consider integrating residual connections into GNN to model arbitrary-order feature interaction. However, as analyzed in Appendix~\ref{app:gnn-res},  incorporating residual connections will significantly elevate the maximum order of feature interaction.

With different orders of feature interactions, we then explicitly combine 
 all nodes embeddings of each node $f_m$ into the updated node embeddings $\hat{\bH}_m$ of $f_m$ by multi-head attention: 
\begin{align*}
\text{attention}(\bQ,\bK,\bV) &= \text{softmax}( {\bQ\bK^\top}/{\sqrt{N^{d}}} )\bV,\\
\text{head}_h &=\text{attention}(\bW_{q,h}\bH_m,\bW_{k,h}\bH_m,\bW_{v,h}\bH_m), \\
\hat{\bH}_m&=
[\text{head}_1;\dots;\text{head}_{N_h}], 
\end{align*}
where $\bH_{m} = [\bh_{m}^{(0)};\ldots;\bh_{m}^{(N_l)}]$ contains $N_l$ row vectors with length $N_d$, and $N_h$ is the number of attention heads. 
Then, we estimate the contribution factor for each of the $N_v+N_u$ features as 
\begin{equation}\label{eq:c}
[c_{1},\ldots,c_{N_v+N_u}] = \text{sigmoid}\left(\text{MLP}_{\bW_{c,1}}([\hat{\bH}_1,\dots,\hat{\bH}_{N_v+N_u}])\right),
\end{equation}
where $\text{MLP}_{\bW_{c,1}}$ is parameterized by $\bW_{c,1}$. 
Finally, 
we predict whether item $v$ and user $u$ interact as 
\begin{align}\label{eq:prediction}
\hat{y} = \sum\nolimits_{m=1}^{N_v+N_u}c_m\cdot\text{MLP}_{\bW_{c,2}}(\hat{\bH}_m), 
\end{align} 
where $\bW_{c,2}$ is a trainable parameter.

\subsection{Learning and Inference}
\label{sec:alg}

To reduce the risk of overfitting when dealing with limited data, 
we introduce a meta learning strategy that optimizes parameters of hypernetworks and GNN across various item CTR prediction tasks, while only adjusting a minimal set of item-specific parameters within each task. 

For simplicity, 
we denote hypernetworks as $\text{hyper}_{\btheta_{\text{hyper}}}$ where $\btheta_{\text{hyper}}=\bW_a$ is the shared trainable parameter. 
Then, 
we 
denote
the GNN as $\text{GNN}_{\btheta_{\text{GNN}},\bphi_i}$, 
where $\btheta_{\text{GNN}}$ represents shared parameters including 
parameters of embedding layers $\{\bW_{e,m}\}_{m=1}^{N_v+N_u}$, 
parameters of GNN layers $\{\bW_g^{(l)}\}_{l=1}^{N_l}$, $\bW_{q,h}$, $\bW_{k,h}$, $\bW_{v,h}$,
parameters of predictor $\bW_{c,1},\bW_{c,2}$, 
and $\bphi_i$ represents item-specific parameters 
\begin{align}\label{eq:phi}
	\bphi_i = \{\text{hyper}_{\btheta_{\text{hyper}}}(v_i), \be_{\text{ID},i}\}, 
\end{align}
which includes item-specific adjacency matrix $\bA^{(1)}$ generated by hypernetworks and the randomized item ID embedding $\be_{\text{ID},i}$ of item $v_i$. 
We target at learning $\btheta^*_{\text{GNN}},\btheta^*_{\text{hyper}}$, which can achieve good cold-start \& warm-up performance on new item $v_i$ by only generating $\bphi_i$ and warming up $\bphi_i$ with gradient descent. 

We optimize \TheName{} w.r.t. the following objective calculated across $N_t$ tasks from $\cT^{\text{old}}$: 
\begin{equation}\label{eq:loss-meta}
	\sum\nolimits_{i}^{N_t}
	\gamma
	\mL_{\cS_i}({\btheta_{\text{GNN}},\bphi_i})
	+
	(1-\gamma)
	\mL_{\cQ_i}({\btheta_{\text{GNN}},\bphi_i'}), 
\end{equation}
where $\gamma$ is a hyperparameter to balance the contribution of two loss terms. 
In particular, 
the first term can represent the performance of cold-start phase~\cite{metaE} as 
the model has not been exposed to the labels in $\cS_i$. 
We compute $\mL_{\cS_i}({\btheta_{\text{GNN}},\bphi_i})$ as
\begin{align}\label{eq:loss-s}
	\lefteqn{\mL_{\cS_i}({\btheta_{\text{GNN}},\bphi_i})}
	\\\notag
	&\equiv {1}/{|\cS_i|}\cdot\sum\nolimits_{(v_i,u_j,y_{ij})\in\cS_i}
	\text{BCE}(y_{ij},\text{GNN}_{\btheta_{\text{GNN}},\bphi_i}(v_i,u_j)), 
\end{align}
where 
$\text{BCE}(y,\hat{y})=-y\log(\hat{y})-(1-y)\log(1-\hat{y})$ is the binary cross entropy. 
The second term in \eqref{eq:loss-meta} represents the performance of warm-up phase after updating $\bphi_i$ by a few new item instances provided in $\cS_i$. 
We compute $\mL_{\cQ_i}({\btheta_{\text{GNN}},\bphi_i'})$ as
\begin{align}\label{eq:loss-q}
	\lefteqn{\mL_{\cQ_i}({\btheta_{\text{GNN}},\bphi_i'})}
	\\\notag
	&\equiv {1}/{|\cQ_i|}\cdot\sum\nolimits_{(v_i,u_j,y_{ij})\in\cQ_i}\text{BCE}(y_{ij},\text{GNN}_{\btheta_{\text{GNN}},\bphi'_i}(v_i,u_j)), 
\end{align}
with 
$\bphi_i'$ obtained by performing gradient descent steps w.r.t \eqref{eq:loss-s}: 
\begin{align}\label{eq:update-specific}
	\bphi_i' = \bphi_i-\alpha_1\nabla_{\bphi_i}\mL_{\cS_i}({\btheta_{\text{GNN}},\bphi_i}),
\end{align}
where $\alpha_1$ is the learning rate. 

Algorithm~\ref{alg:training} summarizes the training procedure of \TheName{}. 

\begin{algorithm}
	\caption{The training procedure of \TheName{}.}
	\label{alg:training}
	\begin{algorithmic}[1] 
		\STATE randomly initialize $\btheta_{\text{hyper}}$ and $\btheta_{\text{GNN}}$;  
		\STATE pretrain $\btheta_{\text{GNN}}$ by old item instances; 
		\FOR{$\cT_i$ in $\cT^{\text{old}}$} 
			\STATE sample $\cS_{i}$ and $\cQ_{i}$ for $\cT_i$;
			\STATE randomly initialize item ID embedding $\be_{\text{ID},i}$ for $\cT_i$;
			\STATE obtain feature embeddings $\be_{1,i},\dots,\be_{N_v+N_u,i}$ by \eqref{eq:emb-layer};
			\STATE generate $\bar{\bA}^{(1)}_i$ of the first GNN layer by \eqref{eq:hyper1};
			\STATE get item-specific parameter $\bphi_i=\{\bar{\bA}^{(1)}_i, \bh_{\text{ID},i}\}$;
			\STATE obtain $\bA^{(l)}_i$ of subsequent GNN layers by \eqref{eq:a-iter}-\eqref{eq:mask}; 
			\STATE obtain feature representations $\bh_{1,i},\dots,\bh_{N_v+N_u,i}$ by \eqref{eq:gnn-update}; 
			\STATE obtain prediction $\hat{y}_{ij}$ by \eqref{eq:prediction} for $(v_i,u_j, y_{i,j})\in\cS_{i}$; 
			\STATE update $\bphi_i$ as $\bphi_i'$ by \eqref{eq:update-specific} with learning rate $\alpha_1$;
			\STATE optimize $\btheta_{\text{hyper}}$ and $\btheta_{\text{GNN}}$ w.r.t. \eqref{eq:loss-meta} by gradient descents with learning rate $\alpha_2$;
		\ENDFOR 		
	\end{algorithmic} 
\end{algorithm}
\begin{algorithm}
	\caption{The testing procedure of \TheName{}.}
	\label{alg:testing}
	\begin{algorithmic}[1] 
		\STATE optimized $\btheta^*_{\text{hyper}}$ and $\btheta^*_{\text{GNN}}$;
		\STATE Consider task $\cT_k$ of a new item $v_k$, given $\cS_{k}$ with a few interaction records of $v_k$ and $\cQ_{k}$ for evaluating CTR prediction performance of $v_k$;
		\IF{$|\cS_{k}|=0$} 
		\STATE randomly initialize item ID embedding $\be_{\text{ID},k}$ for $\cT_k$;
		\STATE obtain feature embeddings $\be_{1,k},\dots,\be_{N_v+N_u,k}$ by \eqref{eq:emb-layer}; 
		\STATE generate $\bar{\bA}^{(1)}_k$ of the first GNN layer by \eqref{eq:hyper1};
		\STATE get item-specific parameter $\bphi_k=\{\bar{\bA}^{(1)}_k, \be_{\text{ID},k}\}$;
		\STATE obtain $\bA^{(l)}_k$ of subsequent GNN layers by \eqref{eq:a-iter}-\eqref{eq:mask}; 
		\STATE obtain feature representations $\bh_{1,k},\dots,\bh_{N_v+N_u,k}$ by \eqref{eq:gnn-update}; 
		\STATE obtain prediction $\hat{y}_{kj}$ by \eqref{eq:prediction} for $(v_k,u_j, y_{k,j})\in\cS_{k}$; 
		\STATE measure performance on $\cQ_{k}$;
		\ELSE
		\STATE update $\bA^{(l)}_k,l \in \{1 \cdots N_l\}$ by \eqref{eq:a-iter} to \eqref{eq:mask};
		\STATE update feature representations $\bh_{1,k},\dots,\bh_{N_v+N_u,k}$ by \eqref{eq:gnn-update}; 
		\STATE update $\bphi_k$ as $\bphi_k'$ by \eqref{eq:update-specific};
		\STATE obtain prediction $\hat{y}_{kj}$ by \eqref{eq:prediction} for $(v_k,u_j, y_{k,j})\in\cS_{k}$; 
		\STATE measure performance on $\cQ_{k}$;
		\ENDIF
	\end{algorithmic} 
\end{algorithm}
By learning from a set of tasks $\cT^{\text{old}}$, the learned $\btheta^*_{\text{GNN}},\btheta^*_{\text{hyper}}$ encode common knowledge. 
Consider task $\cT_k$ of new item $v_k$. 
When $v_k$ has no interaction record, namely $|\cS_{k}|=0$, we obtain its task-specific parameter $\bphi_k$ and test its performance on the test set as cold-start phase performance. Given a few new item instances, we can update $\bphi_k$ to take in the supervised information. 
Once $\bphi_k$ is updated, we measure performance on the test set as warm-up phase performance.  
Algorithm~\ref{alg:testing} describes the testing procedure.

\section{Experiments}
\label{sec:expts}

\subsection{Experimental Settings}
\label{sec:expt-setting}

\paragraph{Datasets.}
We use two benchmark datasets: 
(i) \textbf{MovieLens}
~\cite{harper2015movielens}: a dataset containing 1 million interaction records on MovieLens, whose item features include movie ID, title, year of release, genres and user features include user ID, age, gender, occupation and zip-code; 
and 
(ii) \textbf{Taobao}
~\cite{taobao2018tianchi}: a collection of 26 million ad click records on Taobao, whose  item features include ad ID, position ID, category ID, campaign ID, advertiser ID, brand, price and 
user features include user ID, Micro group ID, cms\_group\_id, gender, age, consumption grade, shopping depth, occupation and city level.	
Following existing works \cite{metaE,CVAR}, 
we binarize the ratings of MovieLens, setting rating smaller than 4 as 0 and the others as 1.

\paragraph{Data Split.}
We adopt the public data split \cite{metaE,MWUF,CVAR}, group items according to their frequency: 
(i) \textbf{old items} which are items appearing in more than $N$ interaction records, where $N=200$ in MovieLens and  $N=2000$ in Taobao; 
and 	
(ii) \textbf{new items} which are items appearing in less than $N$ and larger than $3K$ interaction records, where $K$ is set to 20 and 500 for MovieLens and Taobao respectively. 
To mimic the dynamic process where new items are gradually clicked by more users, 
the interaction records associated with new items are sorted by timestamp. 
We consider
three successive warm-up phases, labeled as A, B, and C, 
each of which involves the introduction of a set of $K$ new interaction records for each item. 
The rest interaction records form testing data for evaluation.

\paragraph{Experiment Pipeline.}
We adopt pipeline of existing works~\cite{metaE,MWUF,CVAR}
to assess how a model adapts to new items over time. 
First, we use old item instances to pretrain the model, and 
directly 
evaluate the model performance on testing data of new items as cold-start phase performance. 
Then, 
we measure how model performs as it learns from a few training data in successive warm-up phases.
In particular, we use the training data in warm-up phases A, B and C to sequentially update the model, and evaluate the performance of corresponding updated models on testing data. 

%
%


\paragraph{Evaluation Metric.}
Following existing works~\cite{CVAR}, the performance is evaluated by 
(i) Area Under the Curve (AUC) \cite{AUC} which represents the degree of separability, 
and 
(ii) F1 score \cite{F1} which is a harmonic mean of the precision and recall.
Both AUC and F1 vary between 0 (worst) and 1 (best). 
\begin{table*}[htbp]
	\caption{Test performance obtained on MovieLens and Taobao. 
		The best results are bolded, the second-best results are underlined.
	}
	\vspace{-5px}
	\label{tab:results}
		\begin{tabular}{c|cc|cc|cc|cc}
		\hline
		\multirow{2}{*}{\textit{MovieLens}} & \multicolumn{2}{c|}{Cold-Start Phase} & \multicolumn{2}{c|}{Warm-Up Phase A} & \multicolumn{2}{c|}{Warm-Up Phase B} & \multicolumn{2}{c}{Warm-Up Phase C} \\	
		& AUC(\%) & F1(\%) & AUC(\%) & F1(\%) & AUC(\%) & F1(\%) & AUC(\%) & F1(\%)\\	
		\hline
		DeepFM & \ms{71.48}{0.15} & \ms{60.96}{0.22} & \ms{75.21}{0.27} & \ms{64.09}{0.01} & \ms{77.70}{0.26} & \ms{66.18}{0.15} & \ms{79.45}{0.19} & \ms{67.81}{0.14}\\
		Wide\&Deep & \ms{69.44}{0.29} & \ms{59.53}{0.31} & \ms{74.82}{0.29} & \ms{64.44}{0.21} & \ms{77.58}{0.25} & \ms{66.82}{0.24} & \ms{79.09}{0.22} & \ms{67.67}{0.27} \\
		AutoInt & \ms{68.64}{0.24} & \ms{59.63}{0.14} & \ms{75.60}{0.31} & \ms{64.93}{0.36} & \ms{77.65}{0.33} & \ms{66.84}{0.42} & \ms{79.20}{0.34} & \ms{67.77}{0.36} \\
		LorentzFM & \ms{68.91}{0.15} & \ms{56.22}{0.27} & \ms{75.35}{0.17} & \ms{62.77}{0.21} & \ms{78.46}{0.08} & \ms{66.23}{0.25} & \ms{79.85}{0.02} & \ms{67.93}{0.08} \\
		AFN & \ms{71.23}{0.42} & \ms{61.76}{0.37} & \ms{74.26}{0.08} & \ms{64.39}{0.09} & \ms{76.19}{0.24} & \ms{65.84}{0.16} & \ms{77.36}{0.35} & \ms{66.71}{0.28} \\
		Fi-GNN & \ms{71.37}{0.05} & \ms{61.46}{0.08} & \ms{74.62}{0.03} & \ms{63.83}{0.12} & \ms{76.83}{0.06} & \ms{65.71}{0.05} & \ms{78.49}{0.05} & \ms{66.74}{0.06} \\
		FinalMLP& \ms{69.51}{0.06} & \ms{60.59}{0.17} & \ms{78.48}{0.12} & \ms{67.34}{0.10} & \ms{78.47}{0.16} & \ms{67.27}{0.07} & \ms{79.07}{0.17} & \ms{68.00}{0.10}\\
		FINAL& \ms{71.64}{0.15} & \ms{61.72}{0.17} & \ms{77.87}{0.13} & \ms{66.99}{0.10} & \ms{77.94}{0.10} & \ms{66.93}{0.14} & \ms{78.29}{0.09} & \ms{67.42}{0.10}\\
		\hline
		DropoutNet& \ms{72.94}{0.17} & \ms{62.43}{0.18} & -&-&-&-&-&-\\
		ALDI & \ms{65.53}{0.13} & \ms{57.47}{0.23} & -&-&-&-&-&- \\
		\hline
		MeLU & - &-& \ms{77.54}{0.06} & \ms{66.71}{0.11} & \ms{79.43}{0.10} & \ms{68.51}{0.05} & \ms{80.26}{0.03} & \ms{68.13}{0.06} \\
		MAMO & -& -& \ms{77.69}{0.10} & \ms{66.92}{0.13} & \ms{79.61}{0.07} & \ms{68.72}{0.04} & \ms{80.37}{0.05} & \ms{68.49}{0.04} \\
		TaNP & - & - & \ms{79.15}{0.10} & \sbms{68.39}{0.14} & \sbms{80.49}{0.17} & \sbms{69.43}{0.15} & \ms{80.71}{0.09} & \ms{69.63}{0.09} \\
		ColdNAS & - & - & \ms{77.45}{0.03} & \ms{67.01}{0.03} & \ms{77.88}{0.12} & \ms{67.25}{0.21} & \ms{78.06}{0.09} & \ms{67.31}{0.11} \\
		\hline
		MetaE & \ms{71.82}{0.70} & \ms{61.76}{0.30} & \sbms{79.53}{0.25} & \ms{67.96}{0.15} & \ms{80.27}{0.09} & \ms{68.31}{0.12} & \ms{80.47}{0.04} & \ms{68.46}{0.12} \\
		CVAR & \sbms{73.58}{0.21} & \ms{63.15}{0.12} & \ms{78.23}{0.10} & \ms{67.03}{0.26} & \ms{80.28}{0.06} & \ms{68.76}{0.12} & \sbms{81.06}{0.04} & \ms{69.33}{0.14} \\
		GME & \ms{71.54}{0.13} & \sbms{64.31}{0.10} & \ms{75.81}{0.20} & \ms{67.50}{0.26} & \ms{78.10}{0.18} & \ms{69.26}{0.20} & \ms{79.15}{0.12} & \sbms{69.95}{0.16} \\
		MWUF & \ms{73.19}{0.66} & \ms{62.61}{0.74} & \ms{78.88}{0.11} & \ms{67.34}{0.22} & \ms{80.26}{0.08} & \ms{68.40}{0.13} & \ms{80.57}{0.05} & \ms{68.66}{0.10} \\	\TheName{} & \bms{75.44}{0.05} & \bms{64.76}{0.15} & \bms{79.92}{0.27} & \bms{68.61}{0.24} & \bms{81.28}{0.21} & \bms{69.71}{0.14} & \bms{81.82}{0.16} & \bms{70.26}{0.14} \\
		 \hline \hline
		\multirow{2}{*}{\textit{Taobao}} & \multicolumn{2}{c|}{Cold-Start Phase} & \multicolumn{2}{c|}{Warm-Up Phase A} & \multicolumn{2}{c|}{Warm-Up Phase B} & \multicolumn{2}{c}{Warm-Up Phase C} \\
		& AUC(\%) & F1(\%) & AUC(\%) & F1(\%) & AUC(\%) & F1(\%) & AUC(\%) & F1(\%)\\
		\hline
		DeepFM & \ms{59.01}{0.84} & \ms{13.47}{0.42} & \ms{60.68}{0.65} & \ms{14.27}{0.20} & \ms{61.51}{0.64} & \ms{14.56}{0.34} & \ms{62.34}{0.54} & \ms{15.00}{0.24}\\
		Wide\&Deep & \ms{59.07}{0.44} & \ms{13.65}{0.06} & \ms{60.92}{0.56} & \ms{14.25}{0.10} & \ms{61.69}{0.51} & \ms{14.56}{0.15} & \ms{62.33}{0.46} & \ms{14.75}{0.13} \\
		AutoInt & \ms{55.69}{1.37} & \ms{12.14}{0.22} & \ms{58.65}{1.26} & \ms{13.61}{0.51} & \ms{59.43}{1.20} & \ms{13.84}{0.36} & \ms{60.07}{1.13} & \ms{14.19}{0.39} \\
		LorentzFM & \ms{56.53}{0.41} & \ms{12.72}{0.04} & \ms{60.83}{0.50} & \ms{14.15}{0.24} & \ms{61.26}{0.47} & \ms{14.31}{0.14} & \ms{61.96}{0.45} & \ms{14.60}{0.14}\\
		AFN & \ms{57.94}{0.99} & \ms{13.28}{0.15} & \ms{58.99}{0.74} & \ms{13.73}{0.20} & \ms{59.81}{0.87} & \ms{13.99}{0.18} & \ms{60.18}{0.69} & \ms{14.18}{0.13} \\
		Fi-GNN & \ms{56.92}{0.08} & \ms{12.79}{0.10} & \ms{60.00}{0.13} & \ms{14.06}{0.18} & \ms{62.09}{0.14} & \sbms{14.82}{0.14} & \ms{62.46}{0.21} & \ms{14.90}{0.05} \\	
		FinalMLP & \sbms{60.64}{0.12} & \ms{13.57}{0.04} & \sbms{63.44}{0.06} & \sbms{14.83}{0.03} & \sbms{63.49}{0.07} & \ms{14.80}{0.03} & \sbms{64.05}{0.02} & \ms{15.05}{0.03}\\
		FINAL & \ms{60.53}{0.24} & \ms{13.63}{0.05} & \ms{63.30}{0.12} & \ms{14.81}{0.06} & \ms{63.35}{0.13} & \ms{14.74}{0.04} & \ms{63.93}{0.12} & \ms{15.01}{0.02}\\
		\hline
		DropoutNet & \ms{60.41}{0.09} & \ms{13.53}{0.02} & -&-&-&-&-&-\\
		ALDI & \ms{50.10}{0.18} & \ms{10.93}{0.05} & -&-&-&-&-&-\\
		\hline
		MeLU & - &- & \ms{61.37}{0.17} & \ms{14.09}{0.17} & \ms{62.48}{0.04} & \ms{14.34}{0.05} & \ms{63.07}{0.07} & \ms{14.64}{0.11} \\
		MAMO & - &- & \ms{61.96}{0.11} & \ms{14.31}{0.09} & \ms{62.52}{0.05} & \ms{14.34}{0.04} & \ms{63.15}{0.12} & \ms{14.78}{0.13} \\
		TaNP & - & - & \ms{55.67}{0.22} & \ms{11.92}{0.31} & \ms{55.85}{0.16} & \ms{12.07}{0.16} & \ms{56.19}{0.09} & \ms{12.08}{0.11} \\
		ColdNAS & - & - & \ms{54.27}{0.07} & \ms{10.89}{0.05} & \ms{54.86}{0.14} & \ms{11.33}{0.13} & \ms{55.01}{0.09} & \ms{11.71}{0.13} \\
		\hline
		MetaE & \ms{59.75}{0.37} & \ms{13.58}{0.06} & \ms{61.19}{0.26} & \ms{14.01}{0.09} & \ms{62.06}{0.31} & \ms{14.41}{0.10} & \ms{62.87}{0.32} & \ms{14.71}{0.07} \\
		CVAR & \ms{60.56}{0.46} & \sbms{13.71}{0.13} & \ms{62.54}{0.19} & \ms{14.38}{0.06} & \ms{63.17}{0.10} & \ms{14.69}{0.05} & \ms{63.95}{0.18} & \sbms{15.09}{0.12} \\
		GME & \ms{60.57}{0.23} & \ms{13.32}{0.33} & \ms{62.55}{0.17} & \ms{13.96}{0.22} & \ms{63.29}{0.05} & \ms{14.39}{0.12} & \ms{63.85}{0.13} & \ms{14.52}{0.08} \\
		MWUF & \ms{59.65}{0.46} & \ms{13.44}{0.15} & \ms{62.08}{0.17} & \ms{14.20}{0.07} & \ms{63.03}{0.13} & \ms{14.63}{0.07} & \ms{63.79}{0.12} & \ms{14.93}{0.06} \\	
		\TheName{} & \bms{61.58}{0.03} & \bms{13.99}{0.05} & \bms{63.56}{0.03} & \bms{15.02}{0.06} & \bms{63.76}{0.02} & \bms{15.15}{0.01} & \bms{64.22}{0.02} & \bms{15.21}{0.02}\\
		 \hline
		\end{tabular}
\end{table*}
\subsection{Performance Comparison}
\label{sec:expts-compare}


We compare the proposed \TheName{}\footnote{Our code is available at \url{https://github.com/LARS-group/EmerG}.} with the following four groups of baselines:
\begin{itemize}[leftmargin=*]
	\item[A] General CTR backbones pretrained using old item instances and fine-tuned by new item instances, including 
	\textbf{DeepFM}~\cite{DeepFM},  
	\textbf{Wide\&Deep}~\cite{WD}, \textbf{AutoInt}~\cite{lorentzfm},
	\textbf{AFN}~\cite{AFN},
	\textbf{Fi-GNN}~\cite{fignn}, recent 
	\textbf{FinalMLP}~\cite{mao2023finalmlp} and 
	\textbf{FINAL}~\cite{zhu2023final}. 
	\item[B] Methods for new items without interaction records, including
	\textbf{DropoutNet}~\cite{volkovs2017dropoutnet} and 	\textbf{ALDI}~\cite{ALDI}.

	\item[C] Methods for new items with a few interaction records, 
	including 
	\textbf{MeLU}~\cite{melu},
	\textbf{MAMO}~\cite{mamo},
	\textbf{TaNP}~\cite{TaNP}, and 
	\textbf{ColdNAS}~\cite{ColdNAS}. 
	These methods cannot incorporate new item instances dynamically. Therefore, 
	to accommodate the training interaction records provided in warm-up phases A, B, and C, we adopt a phased approach: initially, we use $K$ interaction records from phase A as the support set to assess testing performance. Subsequently, we combine $2K$ records from phases A and B as the support set for a second evaluation. Finally, we incorporate $3K$ records from all three warm-up phases—A, B, and C—as the support set to conduct a third assessment of testing performance.
		
	
	\item[D] Methods for emerging items with incremental interaction records, which are the most relevant to ours.  
	Existing works mainly equip general CTR backbones with the ability to generate and warm-up item ID embeddings for new items, including 
	\textbf{MetaE}~\cite{metaE}, 
	\textbf{MWUF}~\cite{MWUF}, 
	\textbf{GME}~\cite{GME}, and
	\textbf{CVAR}~\cite{CVAR}.
	We use the classic DeepFM as the CTR backbone. 
	Results of equipping these methods with other backbones are reported in Appendix~\ref{app:expts-backbone}.
\end{itemize}
We implement the compared methods using public codes of the respective authors. 
More implementation details are provided in Appendix~\ref{app:hyperparameters}. 

\paragraph{Performance for Cold-Start \& Warm-Up Phases.}
Table~\ref{tab:results}  shows the results. 
As shown, 
cold-start methods designed for cold-start \& warm-up phases generally perform better. 
\TheName{} consistently performs the best in all four phases, validating the effectiveness of capturing item-specific feature interaction by hypernetworks. 
Few-shot methods for N-way K-shot settings obtains good performance in warm-up phase A. However, as the number of samples increases, it is unable to achieve greater performance improvement without complete retraining.
General CTR backbones which are fine-tuned using the training sets perform worse, where 
{FinalMLP} performs the best.  
Recall that they randomly initialize item-specific parameters for new items, fine-tuning pretrained models by a small number of new item instances is not enough to obtain good performance.  
Particularly, note that the GNN-based CTR model Fi-GNN which uses item-user-specific feature interaction graphs perform not well. 
This shows that too much freedom is not beneficial to capture feature interaction patterns under cold-start \& warm-up phases. 
While in \TheName{}, we utilize hypernetworks to generate item-specific feature graphs, which is then processed by a GNN with customized message passing mechanism to capture arbitrary-order feature interaction, and optimize parameters by meta learning strategy.  
All these design considerations contributes the best performance obtained by \TheName{}. 
For computational overhead, 
\TheName{} is relatively more efficient in terms of both time and computational resources. 
See Appendix~\ref{app:expts-time} for a detailed comparison. 

\begin{figure}[htbp]
	\vspace{-5px}
	\centering
	\subfigure[MovieLens.]{
		\includegraphics[width=0.225\textwidth]{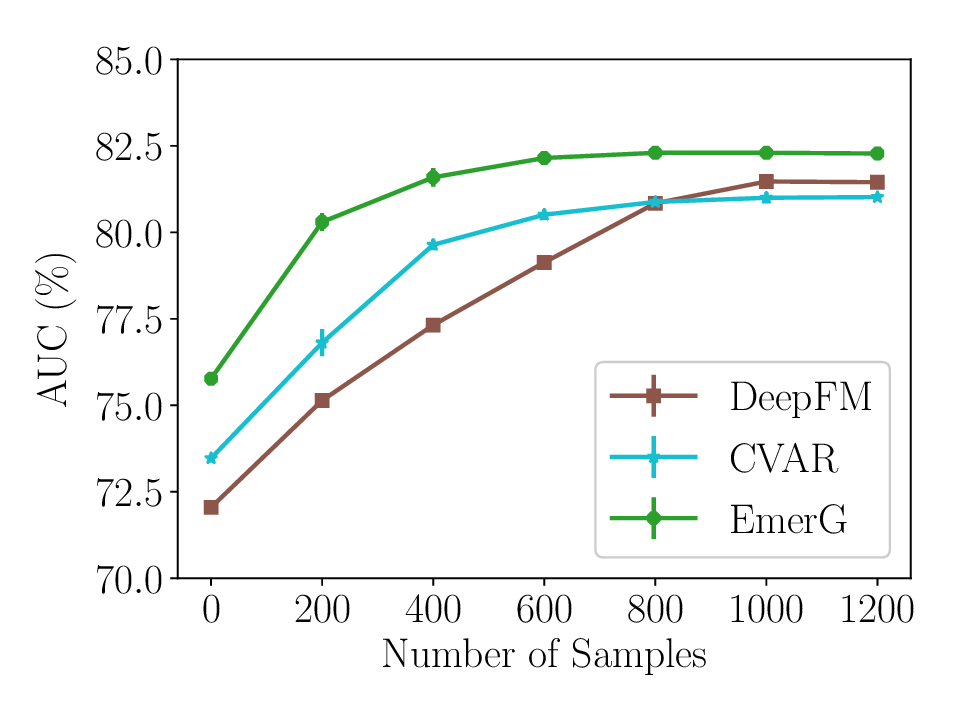}
	}
	\subfigure[Taobao.]{
		\includegraphics[width=0.225\textwidth]{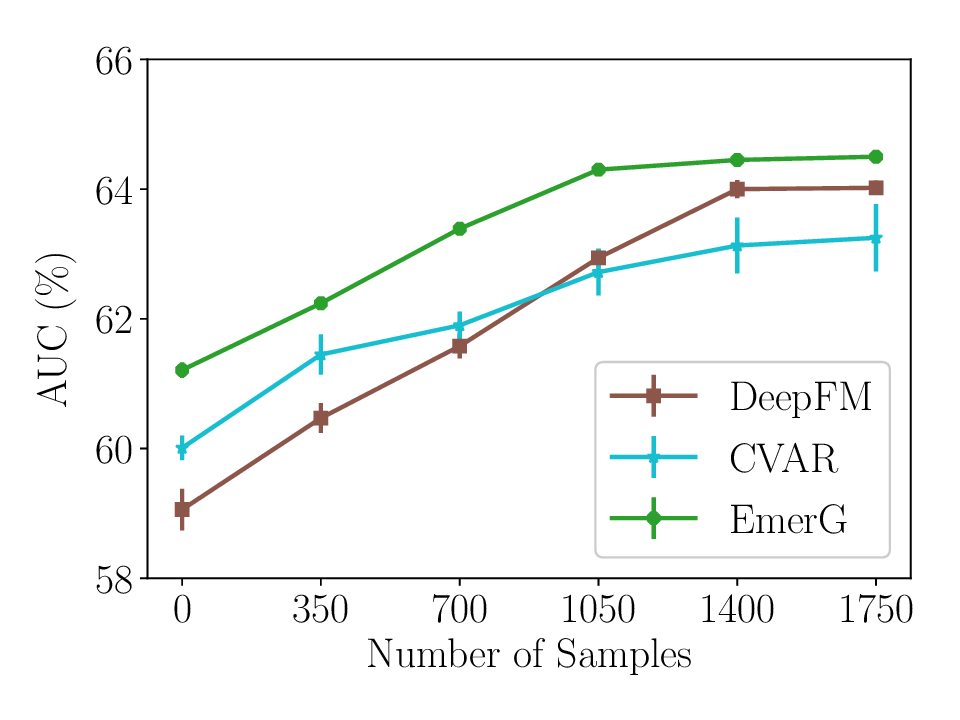}
	}
	\vspace{-5px}
	\caption{Comparing \TheName{} with DeepFM and CVAR given sufficient training samples.}
	\vspace{-5px}
	\label{fig:common-CTR}
\end{figure}

\paragraph{Performance Given Sufficient Training Samples.}
One might question how \TheName{} performs with an abundance of training samples for new items, referred to as the common phase, especially in comparison to traditional CTR backbones. 
 Here, we set aside samples from original testing samples of new items (so the test set is smaller), use them to 
 augment the experiment pipeline with more training samples, and  
 evaluate the performance on the smaller test set. 
We compare \TheName{} with 
baselines which perform the best among CTR backbones and few-shot methods 
in Table~\ref{tab:results}. 
Figure~\ref{fig:common-CTR} shows the testing AUC (\%) with the number of training samples.
Experimental results measured by testing F1 (\%) are similar. 
As shown, all methods get better performance given more training samples. 
The classic CTR backbone DeepFM gradually outperforms 
CVAR which equips DeepFM with additional modules to generate item ID embeddings for new items. 
In contrast, 
\TheName{} consistently performs the best and converges to better performance than the others. 
This validates the effectiveness of our \TheName{} which can nicely capture item-specific  feature interaction at different orders. 

\subsection{Model Analysis}
\subsubsection{Ablation Study}
\label{sec:ablation}

We compare the proposed \TheName{} with the following variants: 
(i) \textbf{w/ random graph} generates the adjacency matrix $\bA^{(1)}$ in \eqref{eq:gnn-update} randomly; 
(ii) \textbf{w/o sparsification} does not apply \eqref{eq:a-sparsify} to sparsify the adjacency matrices; 
(iii) \textbf{w/o mask} does not apply \eqref{eq:mask} to enforce nodes which are disconnected in low-order feature graphs to be disconnected in higher-order feature graphs; 
(iv) \textbf{w/ shared graph} employs global shared adjacency matrices for all items, in contrast to \TheName{}, which utilizes item-specific adjacency matrices;
(v) \textbf{w/o meta} learns both GNN and hypernetworks from old items, without forming tasks and employ a meta-learning strategy; 
and 
(vi) \textbf{w/o inner} directly uses $\bphi_i$ and does not update it to 
$\bphi'_i$ within each task.

\begin{figure}[htbp]
	\centering
	\subfigure[MovieLens.]{
		\includegraphics[width=0.48\textwidth]{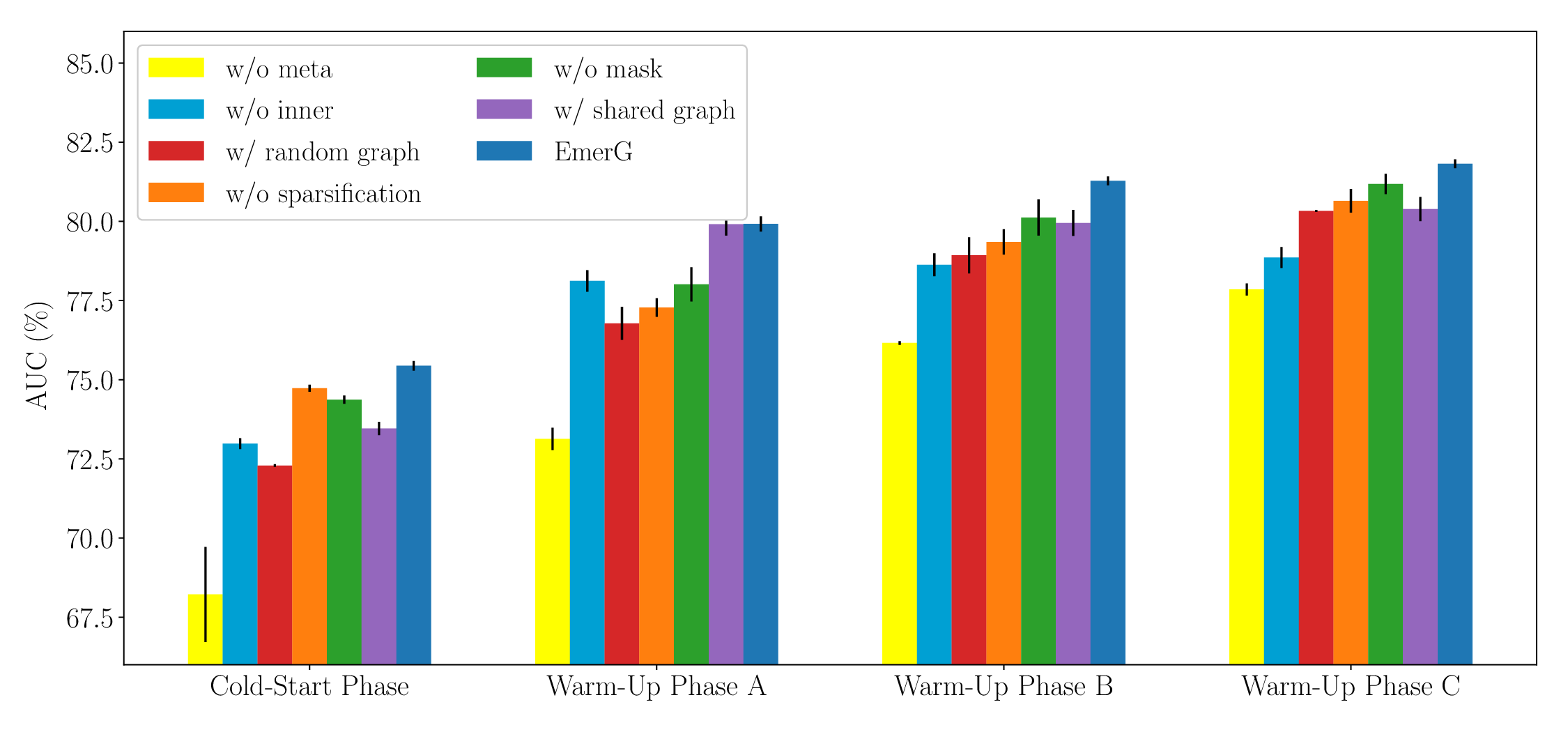}
	}
	\subfigure[Taobao.]{
		\includegraphics[width=0.48\textwidth]{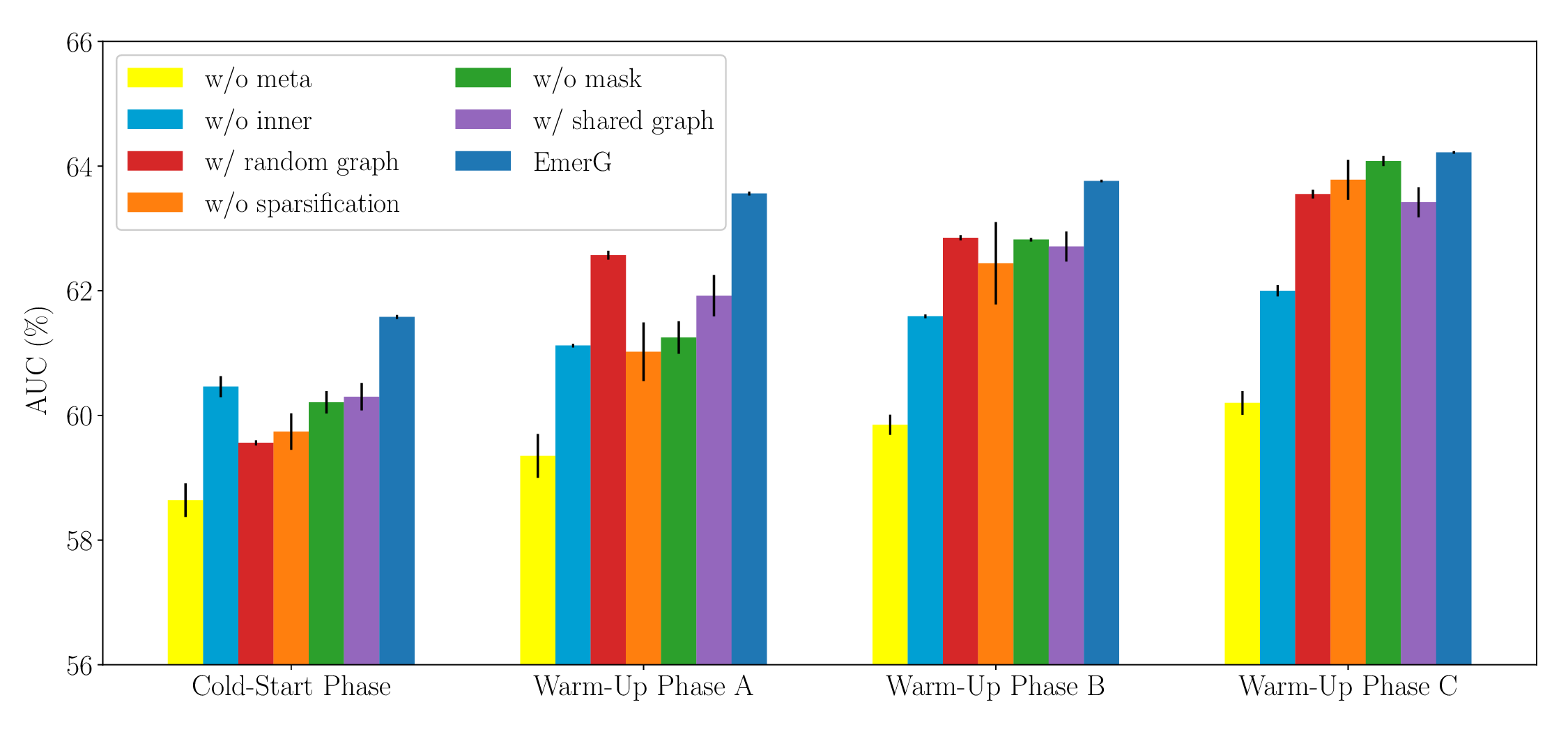}
	}
	\vspace{-5pt}
	\caption{Ablation study on MovieLens and Taobao. }
	\label{fig:ablation}
\end{figure}

Figure \ref{fig:ablation} shows the results. 
As shown, 
``w/ random graph" performs worse than \TheName{} which shows that 
the item-specific feature graphs generated by hypernetworks is meaningful. 
The performance gain of \TheName{} over ``w/o sparsification" shows that a sparse feature graph where only closely-related nodes are connected can let the GNN model concentrate on useful messages. 
Comparing ``w/o mask"  to \TheName{}, the performance drop validates 
our assumption in \eqref{eq:mask}. The mask operation also prevents the adjacency matrices from being too dense, which can be beneficial to prune unnecessary feature interactions and provide better explainability.  
We can also observe that``w/ shared graph" performs worse than \TheName{}.  This validates that using global shared adjacency matrices cannot capture the various feature interaction patterns between different users and items. 
Finally, \TheName{} defeats ``w/o meta" and ``w/o inner", which underscores the necessity of both meta-learning across tasks and inner updates within each task. 

\subsubsection{Effect of Number of GNN Layers} 
As demonstrated in Proposition~\ref{prop:gnn}, we have customized the message passing process of the GNN to ensure that the $l$th layer encapsulates $l$-order feature interactions. Furthermore, we optimize this process by generating adjacency matrices for subsequent GNN layers directly from the initial matrix provided by hypernetworks. This approach not only streamlines the architecture but also facilitates the extension to additional layers, thereby capturing higher-order feature interactions with ease. In this context, we investigate the influence of the number of GNN layers on performance across various datasets.
\begin{figure}[htbp]
	\vspace{-5pt}
	\centering
	\subfigure[MovieLens.]{
		\includegraphics[width=0.225\textwidth]{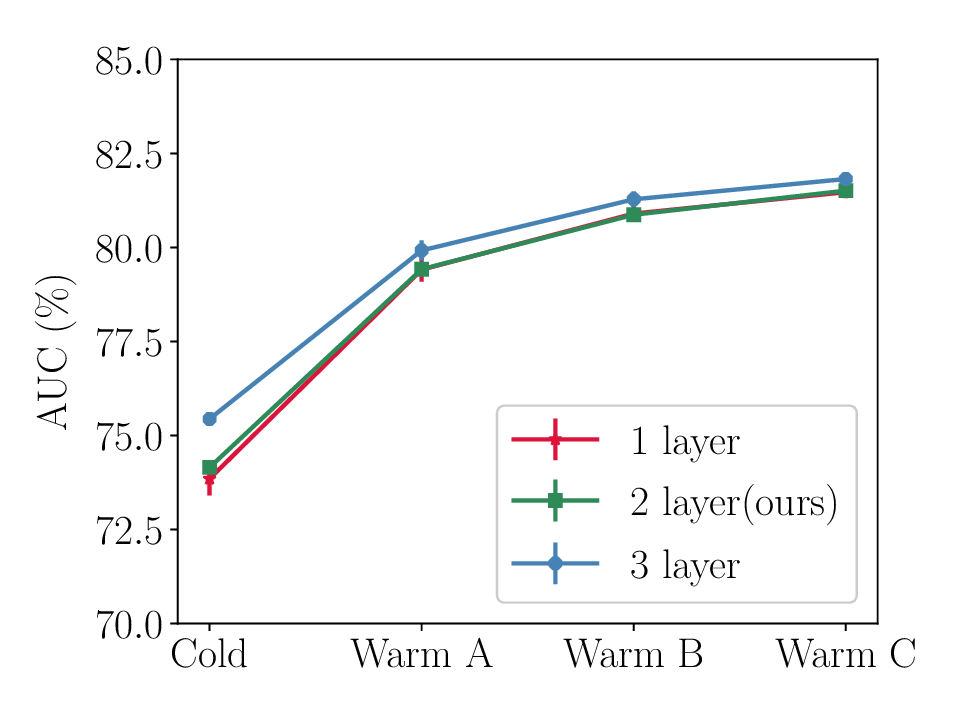}
	}
	\subfigure[Taobao.]{
		\includegraphics[width=0.225\textwidth]{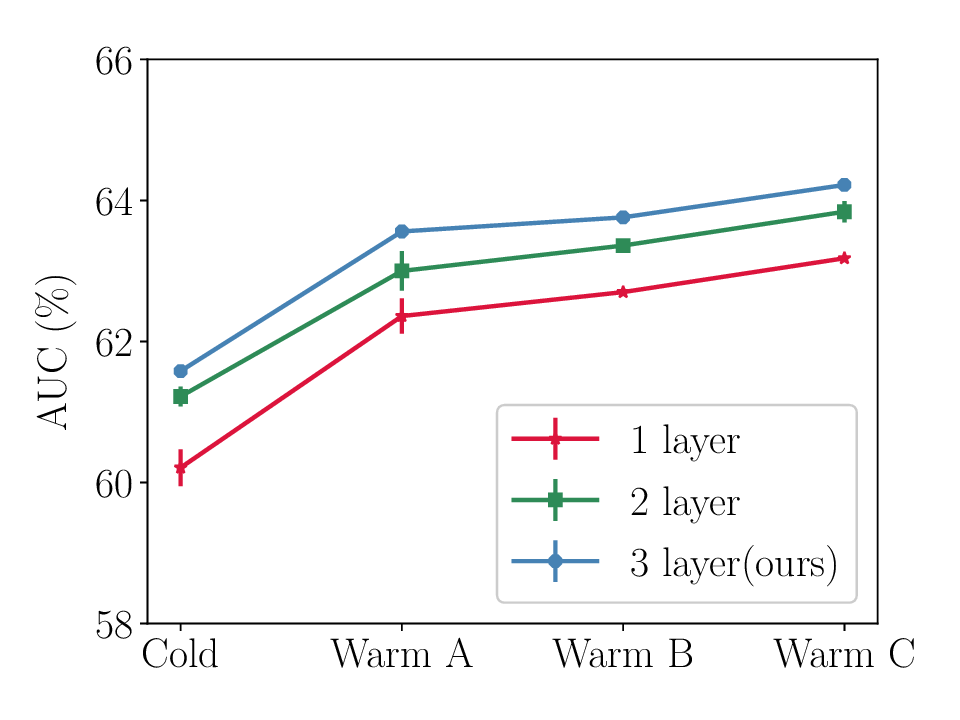}
	}
	\vspace{-5pt}
	\caption{Varying the number of GNN layers in \TheName{}.}
	\label{fig:layeranalysis}
\end{figure}

Figure~\ref{fig:layeranalysis} shows the results. 
As can be seen, \TheName{} with different layer numbers obtain the best performance on different datasets:
\TheName{} with
2 GNN layers performs the best on MovieLens 
while \TheName{} with 3 GNN layers achieves the best performance on Taobao.
This shows that 
different datasets requires different number of GNN layers: 
larger datasets such as Taobao may need higher-order features than smaller ones such as MovieLens. 
By design, \TheName{} can easily meet this requirement. 

	\begin{figure*}[htbp]
	\centering
		\includegraphics[width=1\textwidth]{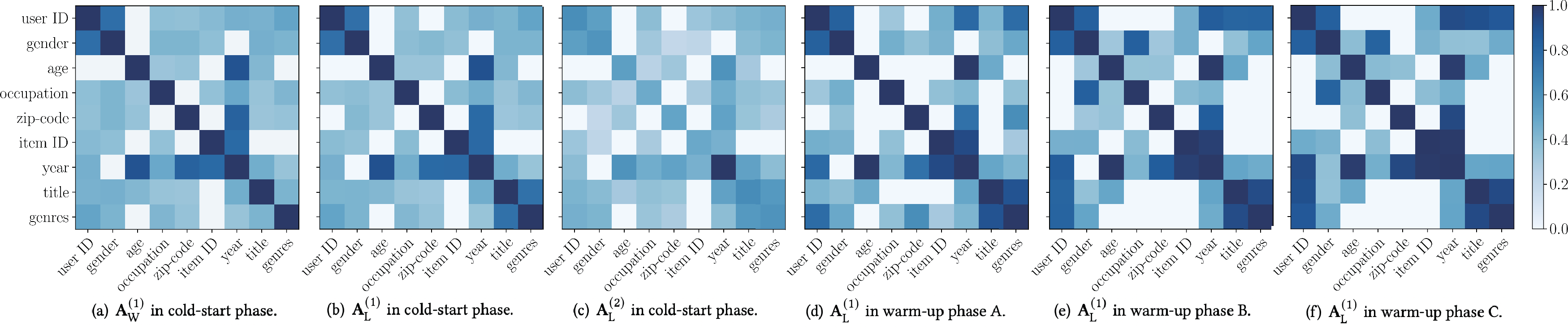}
		\vspace{-5pt}
		\caption{Visualizations of item-specific adjacency matrices of movie \emph{Lawnmower Man 2: Beyond Cyberspace} ($\bA^{(i)}_{\text{W}}$) and movie \emph{Waiting to Exhale} ($\bA^{(i)}_{\text{L}}$)  in MovieLens generated by \TheName{}.
		}
		\vspace{-5pt}
		\label{fig:graphanalysis}
	\end{figure*}
\subsubsection{Different Feature Interaction Functions}
We consider using different feature interaction functions.
Table~\ref{tab:interaction-func} shows the results. 
As shown, using element-wise product performs the best, which is adopted in \TheName{}. 
\begin{table}[htbp]
	\vspace{-5px}
	\centering
	\setlength\tabcolsep{2pt}
	\caption{Test performance in AUC (\%) obtained on MovieLens and Taobao. 
		The best results are bolded.
	}
	\vspace{-5px}
	\label{tab:interaction-func}
		\begin{tabular}{c|c|c|c|c}
			\hline
			{MovieLens}& \multicolumn{1}{c|}{Cold-Start} & \multicolumn{1}{c|}{Warm-Up A} & \multicolumn{1}{c|}{Warm-Up B} & \multicolumn{1}{c}{Warm-Up C} \\ \hline
			$\odot$ & \textbf{75.44} &\textbf{79.92}&\textbf{81.28} &\textbf{81.82}\\ 
			$\max$& 74.14 & 79.43& 81.02& 81.65\\
			$+$&73.64 & 79.31 & 80.77& 81.39\\
			\hline\hline
			{Taobao}& \multicolumn{1}{c|}{Cold-Start} & \multicolumn{1}{c|}{Warm-Up A} & \multicolumn{1}{c|}{Warm-Up B} & \multicolumn{1}{c}{Warm-Up C} \\ \hline
			$\odot$ &\textbf{61.58}& \textbf{63.56} & \textbf{63.76}& \textbf{64.22}  \\ 
			$\max$ &59.98& 62.49 & 62.56 & 63.07  \\ 
			$+$ &59.80& 62.34& 62.26& 62.89  \\ 
			\hline
		\end{tabular}
			\vspace{5px}
\end{table}

\subsection{Case Study}
\label{sec:expts-vis}


Finally, we take 
movie  \emph{Lawnmower Man 2: Beyond Cyberspace}
and movie \emph{Waiting to Exhale} from MovieLens
as new items, and 
visualize their adjacency matrices which record the item-specific feature graphs in Figure~\ref{fig:graphanalysis}. 

As can be seen, 
\TheName{} learns different task-specific feature graphs for different items. 
Comparing Figure~\ref{fig:graphanalysis}(a) with Figure~\ref{fig:graphanalysis}(b), we find that \emph{Lawnmower Man 2: Beyond Cyberspace} has a particularly  important second-order feature interaction $\langle \text{genres}, \text{title}\rangle$. The genre of \emph{Lawnmower Man 2: Beyond Cyberspace} is science fiction while the genre of \emph{Waiting to Exhale} is comedy. For a science fiction, its title often reflects its world view or theme, which is the key for people to judge whether they are interested. As for a comedy work, whether it is interesting or not is often irrelevant to the title.

Besides, 
\TheName{} can capture meaningful higher-order feature interactions. 
As shown, 
both $\langle \text{year}, \text{age}\rangle$  and 
$\langle \text{year}, \text{zip-code}\rangle$ are important second-order feature interactions, they contribute to discovering the third-order feature interaction $\langle \text{year}, \text{age}, \text{zip-code}\rangle$. 
In Figure~\ref{fig:graphanalysis}(c), the relation between nodes of year, age and zip-code all become relatively important although <age, zip-code> is not important in Figure~\ref{fig:graphanalysis}(b).
It is easy to understand that the year of movies determines the age of people who are more likely to watch them. 
For example, elderly people generally prefer watching old movies. Apart from this, location which is indicated by zip-code also plays an important role:  
people in developed areas tend to be more receptive to new things. 
Therefore, area changes may lead to changes in the age of people who like the same movie, which validates 
the efficacy of the learned third-order feature interaction.

We can also observe that 
the item-specific feature graphs generated by our hypernetworks can roughly capture the feature interactions before seeing any training samples of new items. Although 
they 
are continuously optimized using training sets of warm-up phases, the changes are not sharp. 
As can be seen, the second-order feature interaction patterns are similar in Figure~\ref{fig:graphanalysis}(b) and  Figure~\ref{fig:graphanalysis}(d) to Figure~\ref{fig:graphanalysis}(f), with small changes to accommodate the training samples. 
Overall, we conclude that \TheName{} can learn reasonable adjacency matrices to capture item-specific feature interactions at different orders.

\section{Conclusion}
In this study, we underscore the critical role of feature interactions and introduce \TheName{}, a novel solution designed to capture the unique interaction patterns of items, effectively 
handling CTR prediction of newly emerging items with incremental interaction records. 
Our approach leverages hypernetworks to construct item-specific feature graphs, with nodes representing features and edges denoting their interactions, thus enabling the model to discern the intricate interaction patterns that characterize each item. We incorporate a graph neural network (GNN) equipped with a specialized message passing process, crafted to capture feature interactions across all orders, facilitating precise CTR predictions.
To combat overfitting in scenarios with sparse data, we implement a meta-learning strategy that finely tunes parameters of hypernetworks and GNN across various item CTR prediction tasks, necessitating only minimal modifications to item-specific parameters for each task. 
Experimental results on real-world datasets show \TheName{} obtains the state-of-the-art performance on CTR prediction for new items that have no interaction history, a few interactions, or a substantial number of interactions. 
We expect this approach can be used to warm-up cold-start problems in other applications such as drug recommendation in the future. 

\section*{Acknowledgment}
We thank the anonymous reviewers for their valuable comments. 
This work is supported by National Key Research and Development Program of China under Grant 2023YFB2903904 and National Natural Science Foundation of China
under Grant No. 92270106. 

\clearpage
{
\bibliographystyle{ACM-Reference-Format}
\bibliography{ctr}
}
\newpage

\appendix
\section{Message Passing Mechanism}
\subsection{Proof of Proposition~\ref{prop:gnn}}
\label{app:proof}
\begin{proof}
	In \eqref{eq:gnn-update}, node embedding $\bh_{m}^{(l-1)}$ aggregates from multiple first-order node embeddings $\bh_{n}^{(0)}$ to generate $\bh_{m}^{(l)}$. Therefore, 
	\begin{align*}
		\text{order}(\bh_{m}^{(0)}) &= 1,\\
		\text{order}(\bh_{m}^{(l)}) &= \text{order}(\bh_{m}^{(l-1)}) + 1, 
	\end{align*}
	where $\text{order}(\bh_{m}^{(l)})$ represents the order of $\bh_{m}^{(l)}$. 
	We can conclude 
	\begin{align*}
		\text{order}(\bh_{m}^{(l)}) = l+1.
	\end{align*}
	After merging all nodes embedding with multi-head self-attention, all orders of features we can obtain after $l-1$ GNN layers are 
	\begin{align*}
		\text{order}(\hat{\bH}_m)=\{1, 2, ..., l, l+1\}.
	\end{align*}
\end{proof}
\subsection{Comparing with GNN with Residual Connections}
\label{app:gnn-res}
Although GNN with residual connections can model arbitrary-order feature interaction~\cite{fignn}, the maximum order of feature interaction will increase drastically.  To see this, instead of using \eqref{eq:gnn-update}, \eqref{eq:gnn-update-general} can be realized with residual connection as  
\begin{align*}
	\mathbf{h}_{m}^{(l)} = \mathbf{h}_{m}^{(l-1)} + \mathbf{h}_{m}^{(l-1)}\odot{\sum\nolimits_{n = 1}^{N_v+N_u}{[\mathbf{A}^{(l - 1)}_i]_{mn} \mathbf{W}^{(l - 1)}_g\mathbf{h}_{n}^{(l - 1)}}}. 
\end{align*}

As $\mathbf{h}_{m}^{(l-1)}$ and $\mathbf{h}_{n}^{(l-1)}$ have the same maximum order, when we aggregate them via $\odot$, we have: 
\begin{align*}
\text{maxorder}(\mathbf{h}_{m}^{(l)}) = \text{maxorder}(\mathbf{h}_{m}^{(l-1)}) \times 2.
\end{align*} 

In other words, using residue connections will lead to too many noisy high-order feature interactions. Consequently, it is also challenging to determine which feature interactions contribute to the prediction, resulting in low interpretability.

\section{Implementation Details}
\label{app:hyperparameters}

All results are averaged over five runs and are obtained on a 32GB NVIDIA Tesla V100 GPU. 
We use  Adam optimizer~\cite{kingma2014adam}. 
To search for the appropriate hyperparameters, we set aside 20\% old items and form validation set using their samples. 
The performance is then directly evaluated on the validation set of these items, which corresponds to cold-start phase. 
When the hyperparameters are found by grid search, we put back samples of these old items, then follow the experiment pipeline described in Section~\ref{sec:expt-setting} and report the results. 
The hyperparameters and their range used by \TheName{} are summarized in Table~\ref{tab:hyperpara}.
\begin{table}[htbp]
	\caption{Hyperparameters used by \TheName{}. $N'=N_v+N_u$.}
	\centering
	\label{tab:hyperpara}
	\resizebox{0.48\textwidth}{!}{%
	\begin{tabular}{l|l|l|l}
		\hline
		Hyperparameter & Range  &MovieLens & Taobao\\
		\hline	
		number of GNN layers & $[1, 2, 3]$ & 2&3 \\
		$K$ in \eqref{eq:a-sparsify} & $[0, 1,\cdots, N'\cdot N'$ & $\frac{N'\cdot N'}{2}$&$\frac{N'\cdot N'}{2}$ \\
		$\gamma$ in loss function & $[1e-2,\dots,1]$ & 0.1&0.1 \\
		number of heads &$[1,2,\dots,5]$ & 3&3 \\
		embedding dimension &$[10,11,\dots,20]$& 16&16 \\ 
		batch size & $[64, 128, \cdots, 1024]$ & 512&512 \\
		pretraining learning rate & $[1e-4,1e-1]$ & 0.005&0.001 \\
		pretraining epochs &$[1,2,\dots,20]$ &2 &1 \\
		meta-training learning rate $\alpha_2$ &  $[1e-4,1e-1]$& 0.001&0.0001 \\
		meta-training epochs & $[1,2,\dots,20]$ & 11&3 \\
		update learning rate $\alpha_1$ during meta-training &  $[1e-4,1e-1]$ & 0.01&0.001 \\
		update learning rate $\alpha_1$ during warming-up &  $[1e-4,1e-1]$ &0.01 &0.01 \\
		warming-up epochs & $[1,2,\dots,20]$ & 11& 16\\ \hline
	\end{tabular}
}
\end{table}

\section{More Experimental Results} 
\label{app:expts}

\subsection{Comparing with Existing Methods Equipped with Different Backbones}
\label{app:expts-backbone}
In Section~\ref{sec:expts-compare}, we employ DeepFM as the backbone for methods in Group D. Despite this, results in Table~\ref{tab:results} indicate that FinalMLP generally surpasses DeepFM, particularly in the warm-up phases, though not in the cold-start phases. 
Therefore, we further integrate FinalMLP, the top-performing backbone from Group A, into methods in Group D. 
Results are reported in Table~\ref{tab:results2}. 
Notably, FinalMLP, when utilized as a backbone for cold-start methods, does not exceed the performance of configurations using DeepFM. This suggests that more recent CTR backbones cannot effectively handle the CTR prediction of newly emerging items.

\begin{table*}[htbp]
	\vspace{-5px}
	\caption{Comparing EmerG with methods for emerging items with incremental interaction records, using various backbones. 
	Test performance obtained on MovieLens and Taobao. 
	The best results are bolded, the second-best results are underlined.
	}
	\label{tab:results2}
		\begin{tabular}{cc|cc|cc|cc|cc}
			\hline
			\multirow{2}{*}{\textit{MovieLens}}
			& & \multicolumn{2}{c|}{Cold-start Phase} & \multicolumn{2}{c|}{Warm-up Phase A} & \multicolumn{2}{c|}{Warm-up Phase B} & \multicolumn{2}{c}{Warm-up Phase C}\\ 
			& & AUC(\%) & F1(\%) & AUC(\%) & F1(\%) & AUC(\%) & F1(\%) & AUC(\%) & F1(\%) \\	\hline 
			\multirow{2}{*}{{DropoutNet}} & {DeepFM}& \ms{72.94}{0.17} & \ms{62.43}{0.18} & \ms{78.69}{0.01} & \ms{67.17}{0.05} & \ms{78.73}{0.06} & \ms{67.12}{0.04} & \ms{79.28}{0.05} & \ms{67.76}{0.09}\\
			&{FinalMLP} & \ms{72.78}{0.10} & \ms{62.41}{0.16} & \ms{78.55}{0.13} & \ms{67.39}{0.11} & \ms{78.46}{0.07} & \ms{67.18}{0.00} & \ms{78.96}{0.07} & \ms{67.78}{0.10}\\
			\hline
			\multirow{2}{*}{MetaE} & DeepFM& \ms{71.82}{0.70} & \ms{61.76}{0.30} & \sbms{79.53}{0.25} & \sbms{67.96}{0.15} & \ms{80.27}{0.09} & \ms{68.31}{0.12} & \ms{80.47}{0.04} & \ms{68.46}{0.12}\\	
			& {FinalMLP}& \ms{59.50}{4.80} & \ms{53.22}{4.77} & \ms{72.38}{3.44} & \ms{62.11}{3.03} & \ms{75.34}{3.64} & \ms{64.35}{3.19} & \ms{76.98}{3.05} & \ms{65.83}{2.64}\\
			\hline
			\multirow{2}{*}{CVAR} & DeepFM& \sbms{73.58}{0.21} & \ms{63.15}{0.12} & \ms{78.23}{0.10} & \ms{67.03}{0.26} & \sbms{80.28}{0.06} & \ms{68.76}{0.12} & \sbms{81.06}{0.04} & \ms{69.33}{0.14}\\
			&{FinalMLP} & \ms{65.91}{2.58} & \ms{59.02}{1.14} & \ms{77.33}{0.16} & \ms{65.94}{0.45} & \ms{77.90}{0.37} & \ms{66.58}{0.26} & \ms{78.86}{0.32} & \ms{67.62}{0.17}\\
			\hline
			\multirow{2}{*}{GME} &DeepFM & \ms{71.54}{0.13} & \sbms{64.31}{0.10} & \ms{75.81}{0.20} & \ms{67.50}{0.26} & \ms{78.10}{0.18} & \sbms{69.26}{0.20} & \ms{79.15}{0.12} & \sbms{69.95}{0.16}\\
			& {FinalMLP}& \ms{71.56}{0.28} & \ms{63.79}{0.37} & \ms{76.48}{0.36} & \ms{67.81}{0.44} & \ms{78.94}{0.28} & \ms{68.86}{0.34} & \ms{80.04}{0.19} & \ms{69.79}{0.34}\\
			\hline
			\multirow{2}{*}{MWUF} & DeepFM& \ms{73.19}{0.66} & \ms{62.61}{0.74} & \ms{78.88}{0.11} & \ms{67.34}{0.22} & \ms{80.26}{0.08} & \ms{68.40}{0.13} & \ms{80.57}{0.05} & \ms{68.66}{0.10} \\ 
			& {FinalMLP}& \ms{69.02}{0.41} & \ms{59.56}{0.41} & \ms{78.06}{0.37} & \ms{66.88}{0.39} & \ms{79.58}{0.15} & \ms{68.23}{0.07} & \ms{80.12}{0.10} & \ms{68.69}{0.08}\\ 
			\hline
			\multicolumn{2}{c|}{\TheName{}}& \bms{75.44}{0.05} & \bms{64.76}{0.15} & \bms{79.92}{0.27} & \bms{68.61}{0.24} & \bms{81.28}{0.21} & \bms{69.71}{0.14} & \bms{81.82}{0.16} & \bms{70.26}{0.14}\\
			\hline
			\hline
			\multirow{2}{*}{\textit{Taobao}}& & \multicolumn{2}{c|}{Cold-start Phase} & \multicolumn{2}{c|}{Warm-up Phase A} & \multicolumn{2}{c|}{Warm-up Phase B} & \multicolumn{2}{c}{Warm-up Phase C}\\ 
			& & AUC(\%) & F1(\%) & AUC(\%) & F1(\%) & AUC(\%) & F1(\%) & AUC(\%) & F1(\%) \\	\hline 
			\multirow{2}{*}{{DropoutNet}} & {DeepFM}& \ms{60.41}{0.09} & \ms{13.53}{0.02} & \ms{62.48}{0.26} & \ms{14.55}{0.10} & \ms{62.60}{0.26} & \ms{14.68}{0.12} & \ms{63.12}{0.17} & \ms{14.82}{0.08}\\
			&{FinalMLP} & \sbms{60.86}{0.14} & \ms{13.69}{0.09} & \sbms{63.37}{0.08}  & \ms{14.75}{0.03} & \sbms{63.43}{0.01} & \ms{14.84}{0.03} & \sbms{63.98}{0.03} & \ms{15.04}{0.04}\\
			\hline
			\multirow{2}{*}{MetaE} & DeepFM& \ms{59.75}{0.37} & \ms{13.58}{0.06} & \ms{61.19}{0.26} & \ms{14.01}{0.09} & \ms{62.06}{0.31} & \ms{14.41}{0.10} & \ms{62.87}{0.32} & \ms{14.71}{0.07} \\
			& {FinalMLP}& \ms{59.71}{1.00} & \ms{13.12}{0.91} & \ms{62.58}{0.45} & \ms{14.87}{0.13} & \ms{62.68}{0.43} & \ms{14.79}{0.12} & \ms{63.30}{0.44} & \ms{15.13}{0.08}\\
			\hline
			\multirow{2}{*}{CVAR} & DeepFM& \ms{60.56}{0.46} & \ms{13.71}{0.13} & \ms{62.54}{0.19} & \ms{14.38}{0.06} & \ms{63.17}{0.10} & \ms{14.69}{0.05} & \ms{63.95}{0.18} & \ms{15.09}{0.12} \\
			&{FinalMLP} & \ms{60.55}{0.49} & \sbms{13.83}{0.23} & \ms{63.24}{0.47} & \sbms{14.99}{0.22} & \ms{63.25}{1.18} & \sbms{15.03}{0.50} & \ms{63.79}{0.77} & \sbms{15.15}{0.43}\\
			\hline
			\multirow{2}{*}{GME} &DeepFM & \ms{60.57}{0.23} & \ms{13.32}{0.33} & \ms{62.55}{0.17} & \ms{13.96}{0.22} & \ms{63.29}{0.05} & \ms{14.39}{0.12} & \ms{63.85}{0.13} & \ms{14.52}{0.08} \\
			& {FinalMLP}& \ms{60.78}{0.15} & \ms{13.76}{0.06} & \ms{63.10}{0.18} & \ms{14.89}{0.16} & \ms{63.12}{0.04} & \ms{14.81}{0.07} & \ms{63.76}{0.19} & \ms{14.96}{0.14}\\
			\hline
			\multirow{2}{*}{MWUF} & DeepFM& \ms{59.65}{0.46} & \ms{13.44}{0.15} & \ms{62.08}{0.17} & \ms{14.20}{0.07} & \ms{63.03}{0.13} & \ms{14.63}{0.07} & \ms{63.79}{0.12} & \ms{14.93}{0.06} \\	 
			& {FinalMLP}& \ms{60.36}{0.11} & \ms{13.73}{0.08} & \ms{63.26}{0.07} & \ms{14.00}{0.02} & \ms{63.37}{0.22} & \ms{14.79}{0.05} & \ms{63.96}{0.23} & \ms{15.09}{0.05}\\ 
			\hline
			\multicolumn{2}{c|}{\TheName{}}& \bms{61.58}{0.03} & \bms{13.99}{0.05} & \bms{63.56}{0.03} & \bms{15.02}{0.06} & \bms{63.76}{0.02} & \bms{15.15}{0.01} & \bms{64.22}{0.02} & \bms{15.21}{0.02}\\
			\hline
		\end{tabular}
\end{table*}

\subsection{Computational Overhead}
\label{app:expts-time}

Table~\ref{tab:computation} shows 
a detailed comparison of the computational overhead for all compared methods. 
As indicated, \TheName{} demonstrates relatively lower time and space requirements. 

\begin{table}[H]
	\caption{Computational overhead of compared methods on MovieLens. 
	Time is reported as seconds per epoch. 
	}
	\vspace{-10pt}
	\centering
	\label{tab:computation}
		\begin{tabular}{cc|c|c|c}
			\hline
			&& Training Time & Test Time & \# Para. \\
			\hline	
			\multicolumn{2}{c|}{DeepFM} & 831.30 & 43.59 & 0.51 \\
			\multicolumn{2}{c|}{Wide\&Deep} & 725.15 & 45.30 & 0.51 \\
			\multicolumn{2}{c|}{AutoInt} & 877.75 & 48.90 & 0.54 \\
			\multicolumn{2}{c|}{LorentzFM} & 922.75 & 52.21 & 0.51 \\
			\multicolumn{2}{c|}{AFN} & 926.65 & 46.30 & 10.29 \\
			\multicolumn{2}{c|}{Fi-GNN} & 970.11 & 56.11 & 0.53 \\
			\multicolumn{2}{c|}{FinalMLP}  & 1013.55 & 53.30 & 2.30 \\
			\multicolumn{2}{c|}{FINAL} & 944.84 & 49.00 & 1.03 \\\hline
			\multicolumn{2}{c|}{MeLU} & 1123.94 & 52.11 & 0.51 \\
			\multicolumn{2}{c|}{MAMO} & 1299.84 & 52.71 & 0.71 \\
			\multicolumn{2}{c|}{TaNP} & 1089.23 & 31.25 & 0.54 \\
			\multicolumn{2}{c|}{ColdNAS} & 1190.11 & 25.81 & 1.81 \\
			\multicolumn{2}{c|}{ALDI} & 576.40 & 16.40 & 0.51 \\
			\hline
			\multirow{2}{*}{DropoutNet}& DeepFM & 827.20 & 43.20 & 0.51 \\
			&FinalMLP& 1043.23 & 54.26 & 2.3 \\
			\hline
			\multirow{2}{*}{MetaE}& DeepFM & 1235.44 & 44.27 & 0.51 \\
			&FinalMLP& 1319.18 & 52.70 & 2.30 \\
			\hline
			\multirow{2}{*}{CVAR}& DeepFM& 2372.70 & 44.20 & 0.52 \\
			&FinalMLP& 2516.50 & 52.89 & 2.31 \\
			\hline
			\multirow{2}{*}{GME}& DeepFM & 1099.30 & 43.99 & 0.52 \\
			&FinalMLP& 1101.44 & 54.21 & 2.31 \\
			\hline
			\multirow{2}{*}{MWUF}& DeepFM & 1784.95 & 43.67 & 0.52 \\
			&FinalMLP & 2012.60 & 53.44 & 2.31 \\
			\hline
			\multicolumn{2}{c|}{\TheName{}} & 996.26 & 46.10 & 0.82 \\
			\hline
		\end{tabular}
\end{table}

\end{document}